\documentstyle[12pt,aasms4]{article}
\def\m82{M82}
\def\Ab{Abell~}
\def\gi{\mbox {\rm g-i}}

\def\h{\mbox {\rm H}}

\def\msun{\mbox {${\rm ~M_\odot}$}}
\def\zsun{\mbox {${\rm ~Z_{\odot}}$}}

\def\msunyr{\mbox {$~{\rm M_\odot}$~yr$^{-1}$}}
\def\angs{\mbox {~\AA}}
\def\lya{\mbox {Ly$\alpha$~}}
\def\Ha{\mbox {H$\alpha$~}}
\def\Hb{\mbox {H$\beta$~}}

\def\h0{\mbox {~H$_0$}}
\def\q0{\mbox {~q$_0$}}

\def\asec{\ifmmode {'' }\else $''~$\fi}  
\def\amin{\ifmmode {' }\else $'~$\fi}    
\def\arcsper{\ifmmode \rlap.{'' }\else $\rlap{.}'' $\fi} 
\def\arcmper{\ifmmode \rlap.{' }\else $\rlap{.}' $\fi} 
\def\sles{\lower2pt\hbox{$\buildrel {\scriptstyle <}
   \over {\scriptstyle\sim}$}} 
\def\sgreat{\lower2pt\hbox{$\buildrel {\scriptstyle >}
    \over {\scriptstyle\sim}$}} 
\def\kms{~km~s$^{-1}$~}

\def\sb{~ergs~s$^{-1}$~cm$^{-2}$~arcsec$^{-2}$}

\def\flam{~ergs~s$^{-1}$~cm$^{-2}$ \AA$^{-1}$}
\def\cm3{~cm$^{-3}$}
\def\mpc3{~Mpc$^{3}$}

\def\fig{{Figure}}

\def\et{{\rm et\thinspace al.}\ }   

\def\apj{ApJ}
\def\apjs{ApJS}

\def\aj{AJ}
\def\mn{MNRAS}

\def\aa{A\&A}
\def\aasup{A\&AS}

\begin{document}
\title{An Emission-Line Search for  Star-Forming \\
Dwarf Galaxies Toward  Abell 851}

\author{Crystal L. Martin\altaffilmark{1,2,3}, Jennifer Lotz\altaffilmark{2}, 
	Henry C. Ferguson}
	\altaffiltext{1}{Hubble Fellow}
	\altaffiltext{2}{Visiting astronomer Kitt Peak National Observatory
The National Optical Astronomy Observatories are operated by the
Association of Universities for Research in Astronomy, Inc. under
cooperative agreement with the National Science Foundation.}
	\altaffiltext{3}{Present Address: Caltech 105-24, Pasadena, CA 91125}
\affil{Space Telescope Science Institute, 3700 San Martin Drive, Baltimore, 
MD 21218}

\begin{abstract}

We present deep images of the redshift $z = 0.41$ galaxy cluster \Ab851
(Cl~0939+4713)
including a narrow bandwidth image of the redshifted [OII]~$\lambda\lambda 
3726, 29$ line emission.  The [OII] doublet is the most accessible tracer of 
star formation out to redshifts near unity, and we use it to identify 
candidates for star-forming galaxies.  The galaxies selected lie within a 
projected clustercentric distance of 2~Mpc and have line-of-sight velocities 
with respect to the cluster in the range between $\pm\ 1460$\kms.  The sample 
is complete to magnitude  $m_{5129}(AB) = 24.0$ for galaxies with emission-line
equivalent widths greater than about 11\angs\ in the observed frame.  The 
sample is 
contaminated by foreground galaxies with a steep 4000\angs\ break across the 
filters and possibly \lya\ emitters at z=3.31, but such galaxies can be 
identified using color, brightness, and equivalent-width criteria. Excluding
the likely interlopers, the number density of star-forming galaxies 
near Cl~0939+4713 is still at least three to four 
times higher than that of a similar 
column through the field at z=0.4. Galaxies in the cluster sample have lower
[OII] equivalent widths than similarly selected field galaxies at $z \sim 0.4$
however, and we suggest that while the density of galaxies is higher in
the cluster the star-formation activity in individual galaxies is  supressed
relative to their counterparts in the field. The relatively small enhancement in
the surface density of the [OII]-selected population toward the cluster core
leads us to believe the [OII]-selected galaxies reside in the outer regions
of the cluster.  We identify a population of starburst galaxies among the 
[OII]-emitters based on their blue g-i color.  The majority of the 
[OII]-selected galaxies appear to be related to the normal spiral galaxy 
population, but the bursting population is dominated by intrinsically faint 
galaxies of below average size.  We suggest that this population 
is composed primarily of dwarf galaxies and discuss their relation to the dwarf 
galaxies in present day clusters.

\end{abstract}

\section{Introduction}
Cosmological models for the hierarchical growth of structure predict
an abundance of low mass galaxy halos.  The properties of the dwarf
galaxies residing in the halos that survive to the present epoch are
extremely sensitive to their star formation histories.  A better 
understanding of how the star formation rate (SFR) evolves in dwarf galaxies
is needed to understand their role in the assembly of galaxies.

Most of the dwarf galaxies in the local universe, at least in terms of numbers, 
are gas-poor galaxies with smooth light distributions whose surface brightness 
declines exponentially with radius (Ferguson \& Binggeli 1994).  We refer to 
these galaxies as dwarf elliptical galaxies  or dE's. This broad
class includes the dwarf spheroidal galaxies of the Local Group but not
M32-like dwarfs which  populate a different locus in the surface brightness
vs. magnitude plane (Ferguson \& Binggeli 1994, 
Mateo 1999).  The nature of the star-forming progenitors of dE's, which
were  not necessarily dwarf galaxies, remains highly controversial.
Two factors strongly suggest environment has played a significant
role in shaping their star formation history.  First, the number of 
dE's per giant galaxy increases with the richness of the environment 
(Ferguson \& Sandage 1991). Second, the duration of the inferred
quiescient period between bursts of star formation in many dwarf spheroidals
is of order 1 Gyr (Smecker-Hane \et 1994; Grebel 1997). Starburst-driven 
winds and superbubbles lift gas out of the disks of dwarf galaxies
on timescales of 10 to 100 Myr (Martin 1998; Mac Low \& Ferrara 1999).
Much of this gas is predicted to fall back down on the disk over
comparable time periods, so a higher burst frequency is expected 
from self-regulated star formation. The orbital timescales of satellite 
galaxies and the frequency of tidal interactions between field galaxies offer a 
more natural match to the measured burst interval (Moore \et 1996).

It remains unclear when a typical dE galaxy formed the bulk of its stars, but
some of the progenitors must have been forming stars as recently as $z \sim 0.4$.
The isochrones fitted to the color -- magnitude diagrams of
many local group dwarf spheroidal galaxies provide direct evidence for starbursts a 
few Gyrs ago (Smecker-Hane \et 1994; Grebel 1997).  Although
the color - magnitude relation of the dE's in nearby clusters
has typically been interpreted as evidence that the bulk of their stars formed
very early, an intermediate age population in these 
galaxies would be difficult to detect or distinguish (cf. Thuan 1985; Bothun \et 1986).
But do clusters at moderate redshifts even contain star-forming dwarf galaxies?
And, if so, is this population very different from the star-forming population
in present epoch clusters? Studying the dynamics of such a population could 
help characterize the gas removal mechanism.

To date, the dwarf galaxy population in moderate redshift clusters has only been 
detected as a statistical excess of faint galaxies relative to the field 
(Driver \et 1994; Wilson \et 1997; Trentham 1997b; Trentham 1998).
Dressler \et (1999) and Dressler \& Gunn (1982) describe
extensive spectroscopic surveys of \Ab 851 (Cl 0939+4713), but their completeness limits 
are not yet deep enough to describe the dwarf galaxy population well.   
Dressler \et (1999) recovered cluster members in one out of every 
two spectra to $m \approx 22$, but the efficiency of this direct approach will 
decline toward fainter flux levels.  (The number of background galaxies rises 
more steeply with magnitude than the number counts of cluster galaxies are
likely to increase with magnitude.) Once cluster members have been identified
at $V = 22$ to $V =24$, spectroscopy with 10~m class telescopes can be used
to study their properties.  For example, Koo \et (1997) find the 
emission-line widths of three (possibly four) of the six compact galaxies 
observed so far in Cl0024+1654 are quite narrow; and Kobulnicky \et (1999)
report measurements of the O/H abundance ratios for two of these galaxies.

Since a large fraction of cluster galaxies at $z \sim 0.4$ have experienced 
recent star formation (Butcher \& Oemler 1978; Dressler \& Gunn 1982),
emission-line selection would appear to be an efficient means of 
identifying the faint cluster members.  To find these emission-line galaxies 
over regions much larger than 
the cluster core, we have undertaken a deep narrowband imaging survey of 
moderate-redshift clusters. One filter isolates redshifted 
[OII] $\lambda \lambda 3726,3729$ emission line radiation from galaxies 
near the mean cluster redshift, and the continuum is measured in a second
deep image.  The main advantage of the [OII] line is that it remains in
the optical region out to redshifts of $z \sim 1$.
The major limitation of this method is the limited accuracy of [OII]
as a tracer of the number of young massive stars.  Fortunately, the [OII]
flux from nearby galaxies correlates with their \Ha luminosity, which is
a direct measure of the SFR (Gallagher \et 1989; Kennicutt 1992; 
Loveday \et 1999).  The scatter in this relation and uncertainty regarding
the amount of attenuation from dust in the galaxies limit how accurately
we can estimate SFRs for the objects we find, but the method does identify
a much larger sample of candidates for follow-up studies 
than has previously been available.

In this paper, we discuss our selection technique and present results for
the cluster \Ab851, also known as Cl 0939+4713. This extraordinarily rich cluster
contains 160 members to R = 22.5 (Belloni \& Roeser 1996), and its X-ray
luminosity approaches that of the Coma cluster (Sadat 1998).  Dressler \et (1999)
present spectra for 71 cluster members and derive a mean redshift of $z = 0.4060$
and cluster velocity dispersion of $\sigma = 1260$\kms.  The cluster core contains a
 high fraction of blue galaxies, and many of these have irregular morphologies 
(Dressler \et 1994; Smail \et 1997). Our observations of this cluster and the 
resulting sensitivity to line emission are described in \S2. Sample contamination 
from other spectral features is discussed in \S3\ prior to comparing the
cluster galaxies to the [OII]-selected population of field galaxies.  In \S4 we discuss 
 the star formation histories of the [OII] selected galaxies, argue that some
of these galaxies are dwarf galaxies in a starburst phase, and speculate about the
nature of their remnants. Our conclusions are summarized in \S5.

\section{Data}

\subsection{Observations}  \label{sec:observations}

Images of \Ab851 were obtained using the T2KB CCD and prime focus 
camera on the KPNO 4m telescope on the nights of 1998 January 23-26.
Filter W022 was chosen as the redshifted [OII] filter based on
the redshift and cluster velocity dispersion published by Dressler \& Gunn 
(1992) which are 0.4067 and 860\kms, respectively.  The revised redshift
from Dressler \et (1999) still places the cluster in the center of this
bandpass, but the larger velocity dispersion means
our survey is less complete than we originally intended.\footnote{Note 
	that the velocity dispersion listed in  Dressler \& Gunn (1992), 
	860\kms, should be 1060\kms\ for the data presented there.}
Figure~\ref{fig:filters} 
shows the effective transmission curve of W022 in the f/3.14 prime focus beam.
Using the revised value $\sigma = 1260$\kms, the peculiar velocity of a galaxy
with respect to the cluster must be between $-1.17\sigma$ and $1.15\sigma$ for
the [OII] line to be redshifted into the region of filter W022 where the transmission
is greater than 70\%.  Roughly 75\% of the virialized cluster galaxies should 
be detected by our survey.  Note, that this redshift interval samples a
much larger proper distance than the angular diameter distance subtended
by our field, which is $\sim 4$~Mpc on a side at z= 0.4.\footnote{
	All calculations in the paper assume
	\h0 = 70 km s$^{-1}$~Mpc$^{-1}$; \q0= 0.1.}  
We chose a continuum filter W021 with a central
wavelength at $\lambda 5129$\angs, which is 116\angs\ blueward of the center
of the on-band filter.

The individual frames were bias subtracted, corrected for pixel-to-pixel
sensitivity variations, and corrected for the large-scale illumination
pattern.  An iterative cosmic-ray rejection algorithm, CRREJ in IRAF, was then 
used to combine the photometric frames in each filter. 
The telescope was offset 60\asec\ between pairs of exposures in a pattern
that makes the exposure time constant over most of the 14\farcm3 field of view. 
The total exposure time in the photometric on-band and off-band frames
was 13,200~s and 10,800~s, respectively. The photometric images were flux 
calibrated with observations of Kitt Peak spectrophotometric standards 
(Massey \et 1988).  More images were taken under good, but not photometric, 
conditions bringing the total exposure time in each band to 18,000~s. 
The sum of these deeper images is shown in Figure~\ref{fig:sum}. The region
shown is the central 1/9th of our field, and the remainder of the field can be viewed at http://astro.caltech.edu/~clm/dwoii.html.
The count rates were a few percent lower than in the photometric 
images, so we scaled the sensitivity by the mean count ratio of the 1000 
brightest objects. The standard deviation of this distribution of flux
ratios was added in quadrature to the  uncertainty in the original flux 
calibration, and this value was used as the uncertainty in the flux 
calibration of the deep images.  The root-mean-square (RMS) variations in 
the sky brightness is $\sigma_{sky} = 
2.11 \times 10^{-18}$\sb\ and $4.1 \times10^{-18}$\sb\ 
in the on-band and off-band images, respectively.  
The seeing, as measured with the IRAF task IMEXAMINE,  was 1\farcs2 FWHM
(full width at half maximum intensity).

Broadband images were also obtained on the night of 1998 January 26 through 
the standard Sloan g and i filters (Fukugita \et 1996).  The total exposure
time in each band was 1 hour. Conditions were partly cloudy and the
net image quality was 2\farcs2 FWHM in g and 2\farcs5 FWHM in i.  The reduction
of these images was similar to that of the narrowband images.
An acceptable g-band calibration, root-mean-square (RMS) scatter of 0.22 mag.,
was obtained from published Gunn magnitudes (Dressler \& Gunn 1992) 
and the transformations described by Jorgenson (1994) and Fukugita \et (1996). 
Our i-band photometry was calibrated using new measurements 
of calibrated WFPC2 F814W frames that were obtained from the HST archive.
We used the typical magnitude differences between these two sets of photometry,
an RMS value of 0.17 mag., to estimate the uncertainty level of the flux calibration.

\subsection{Detection and Photometry}

The combined W022 and W021 image was convolved with a gaussian of FWHM=4.0 
pixels and searched for groups of three or more connected pixels each with 
signal greater than $1.5 \sigma_{sky}$ using SExtractor (Bertin \& Arnouts 
1996).  A total of 5842 objects were detected, and nearly all detections 
included 5 or more pixels. Only one source was detected on the inverse image 
using the same parameters, so the false detection rate is believed to be
quite low.  Sources found within $\sim 10$ pixels of the edge of the field
were omitted from the catalogs reducing the total area surveyed to 
181 arcmin$^2$.  A background map was generated for each frame by fitting 
a smooth surface to a mesh of the local median sky values.
Thirty pixel by 30 pixel cells proved large enough to remove objects and
provided sufficient resolution to model residual scattered light.

Apertures were defined at a constant isophotal brightness on the summed 
narrowband images. After convolving the g and i images to the FWHM image 
quality of the i image, isophotal magnitudes were tabulated in g and i as well
as W021 and W022 using SExtractor.
Errors in the mean background are not included in the magnitude error
returned by SExtractor, which includes only the pixel-to-pixel variations
in the sky level and poisson noise.  
Comparison of a range of acceptable background maps showed only 
a few pixels with discrepancies of several counts.  The error introduced by the 
uncertainty in the mean background level is less than 10\% of the 
error from the pixel-to-pixel fluctuations, $\sqrt{A\sigma_{sky}^2}$, in all 
but a couple of the largest galaxies.

Figure~\ref{fig:stars} illustrates how stars were
distinguished from small galaxies.
The distribution of source sizes over the full range of source brightness  
is clearly bi-modal. 
 We classified
89 objects as stars based on the concentration of their flux and a model fitted
to the point-spread-function of the image.
They are represented by circles in Figure~\ref{fig:stars}.
  These objects are unresolved on the ground-based images,
and the two that fall on the HST field are clearly stellar. 
This distinction becomes ambiguous fainter 
than $F_{\lambda} = 10^{-17}$\flam, or m = 21.5 AB magnitudes, but 
the stars at those magnitudes are greatly outnumbered by galaxies.
The areal density of bright stars is consistent with the measurements of Jones
\et (1991).

\subsection{Sensitivity to Emission Lines} \label{sec:sfr}

We refer to excess on-band flux (relative to the off-band flux) as {\em line flux}
since [OII] emission at the cluster redshift will be redshifted into the
W022 filter. We  will discuss other spectral features that might produce an on-band excess 
in the next section.
Scaling the flux in the off band to the effective width of the on-band,
the transmitted line flux is given by 
$$T(\lambda_l) F_l = F_{ON} - F_{OFF} {\int^{ON} T(\lambda) d\lambda} /
{\int^{OFF} T(\lambda) d\lambda}, $$ where the error on $T(\lambda_l) F_l$ 
includes the photometric and calibration errors.  The true line flux, $F_l$, 
depends on the value of the filter transmission, $T(\lambda_l)$, at the 
wavelength of the redshifted emission from a particular galaxy.  
The mean filter transmission is 78\% over the bandpass with $T_{\lambda} > 0.70$.
The measurement error includes shot noise from the galaxy plus sky, the
pixel-to-pixel fluctuations in the background level, and the uncertainty
in the flux calibration.
For each detected galaxy, Figure~\ref{fig:fsig} shows the significance of 
the on-band excess, $T F_{\l}$,  with respect to the standard deviation of the 
measurement error,  $\sigma$.  Most of the 5560 galaxies detected cluster 
about the line where this signal-to-noise ratio is zero.  They
show no excess emission in the on-band, and those near the cluster redshift
do not have significant [OII] line emission.

Galaxy detection depends on the total line plus continuum flux in the 
co-added bandpasses, but the certainty with which we can detect an emission 
line depends  on the contrast between the line and the continuum. The observed
 equivalent width,  $EW_o \equiv T F_{l} / F_{\lambda}$, describes this 
contrast. (Recall that the galaxy's rest frame spectrum, $EW_e$, is a  
factor of $1+z$ higher.)  In Figure~\ref{fig:fsig}, the 
solid lines illustrate the signal-to-noise (S/N) expected for galaxies with 
equivalent widths of 300\angs, 60\angs, and 11\angs. At fixed equivalent 
width and  angular size, the S/N ratio increases as the square root of the 
continuum flux up to the brightness where the uncertainty in the flux 
calibration begins to dominate the measurement error.  For a bright galaxy, 
the significance level of an on-band flux excess depends almost entirely on 
the line equivalent width.  For example, Figure~\ref{fig:fsig} shows that
a cut at $S/N = 4$ is equivalent to an equivalent width cut at
11\angs.  A cut at $S/N = 3$, dotted line, stretches the equivalent width limit to 8\angs.
This limiting equivalent width describes our selection function
for galaxies brighter than $F_{\lambda} = 10^{-18}$\flam\ which
corresponds to $m_{5129}(AB) = 24.0$.  Fainter than this completeness limit, 
the limiting equivalent width is higher. The size of a faint galaxy 
also influences the significance of a line detection.  As illustrated by 
curve {\it b} and curve {\it c} in Figure~\ref{fig:fsig}, compact galaxies 
are detected at  slightly higher significance at a given equivalent width and 
magnitude.  

The S/N ratio of the on-band flux excess exceeds three for 371 galaxies, 
and these objects define our {\it on-band-selected sample} of emission-line 
candidates. A more conservative cut, at say $S/N = 4$, would reduce the number
of emission-line candidates to 258.  The preferred cut depends upon how the 
sample is to be used, so we provide the entire $S/N \ge\ 3$ catalog to allow 
other researchers to choose a selection level appropriate to their goal. Based
on the relative distribution of positive and negative detections in
\fig~\ref{fig:fsig}, we felt the cut at $S/N = 3$ was appropriate for
spectroscopic follow-up.  The population of 106 objects with a significant, 
i.e. $S/N \ge\ 3 $, flux excess in the off-band defines our {\it off-band 
selected sample} of galaxies.

Table~1 lists 33 on-band-selected galaxies with colors bluer than
$\gi\ = 1.0$. These candidates for [OII]-emission are ordered by
the significance of their on-band flux excess, $T_{\lambda} F_l / \sigma$,
from highest to lowest.  Similar tables for galaxies with redder colors are
available at http://astro.caltech.edu/~clm/dwoii.html. 
The catalog number (col. 1) is related to a galaxy's position in our 
field (cols. 2 and 3).  The origin is at the southeast corner of our frame.  
We fitted a coordinate transformation to an image extracted from the 
Digitized Sky Survey to derive the absolute positions shown in columns 4 
and 5. The root mean square deviation of the fit residuals was 0\farcs2. 
The on-band flux excess is given in col. 6, and col.~7 lists the continuum 
flux. Cols. 8 and 9 give the broad band flux and color.  The nature of 
these objects  is discussed in \S~\ref{sec:results}.

\subsection{Star Formation Rate Estimates}

To gain some insight into the SFR's of the [OII]-selected galaxies,
we must adopt an empirical relation between the measured [OII] and \Ha 
fluxes of the galaxies.
The \Ha luminosity of a galaxy provides an excellent measure of the
massive SFR because the recombination rate of the interstellar gas 
directly reflects the ionization rate.  Collisions between the same
free electrons and singly ionized oxygen atoms produce the excited states
that decay via the [OII] $\lambda\lambda 3726,29$ transitions.  It is
therefore not surprising that the [OII] and \Ha fluxes are correlated
(e.g. Gallagher \et 1989; Kennicutt 1992a, Loveday \et 1999).
The empirical relation among the bluest local galaxies 
shows a scatter of a factor of three about the mean [OII] versus \Hb
relation, $F_{[OII]} = 3.2 F_{H\beta}$ (Gallagher \et 1989). In the
absence of reddening, the equivalent [OII] -- \Ha relation would be
$F_{[OII]} = 1.17 F_{H\alpha}$ assuming a nebular temperature of $10^4$~K
and Case~B recombination (Osterbrock 1989).  Throughout this paper  
we adopt $A_B = 1.0$ as a representative extinction correction for dwarf 
irregular galaxies (e.g. Martin 1997), and the reddened \Ha\ flux is 
predicted to be $1.17 F_{[OII]}$. This approximation will systematically
underestimate the SFR's of spiral galaxies because their fitted $F_{[OII]} / 
F_{H\alpha}$ ratio is slightly lower (Kennicutt 1992a; Loveday \et 1999).

Our estimate of the \Ha flux needs to be corrected for dust attenuation
before the \Ha luminosity, $L_{H\alpha}$ is used to estimate the SFR.
The variation in the dust attenuation from galaxy to galaxy is probably
at least as large as the differences in the intrinsic [OII]/\Ha ratio.
After the extinction correction, $A_B = 1.0$~mag ($A_{H\alpha} = 
0.59$~mag), is applied, the extinction-corrected \Ha fluxes are 2.0 
times larger than the measured, reddened [OII] fluxes.  Choosing
an extinction correction appropriate for dwarf irregular galaxies will
systematically underestimate the attenuation (and SFR) in more luminous 
galaxies.  The SFR calibration derived by Kennicutt \et (1994)
assumes stellar masses from  0.1 to 100\msun\  populate the K83 IMF 
(Kennicutt 1983) and evolve along evolutionary tracks from Schaller et al. 
(1992). Applying their calibration, $R = L_{H\alpha} / 1.36 \times 
10^{41}$~ergs~s$^{-1}$\msunyr, to the galaxies with blue colors in 
Table~1, we find that half of these galaxies have star formations rates
less than 0.2\msunyr.  The median SFR in the galaxies with intermediate
colors is more than twice as large.

\subsection{Incompleteness} \label{sec:incomplete}

We ran Monte Carlo simulations in which artificial images were created with 
a similar number of galaxies and the same noise as the data. We extracted 
sources from these images with SExtractor using the parameter settings that 
had been tuned to the real data.  Comparisons between the resulting catalogs 
and input lists revealed  which galaxies were missed.  By repeatedly adding 
several thousand galaxies to the images at a time, we  measured the detected 
fraction of galaxies at each point across a grid of line fluxes and continuum 
fluxes.  We used this grid to assign a detection probability, $p_{ij}$, to each 
galaxy.  The galaxies that we added to the images were viewed at random 
inclinations, but all had exponential surface brightness profiles with a 
scale length of 1~kpc (0\farcs2).
The detection probabilities  were not senstive to the scale length adopted.

Contours of 85\% and 3\% completeness are superposed on the on-band-selected
sample in \fig~\ref{fig:on_l}. Galaxies with the same equivalent width lie 
along diagonal lines in this diagram.  Bright galaxies, on the right half of 
the diagram, are shown down to an equivalent width  $EW_o \sim 8\angs$.  
The simulations do a good job of predicting the  break in the data near
$\log F_{\lambda} = -18$, i.e $m_{5129}(AB) = 24.0$,   where
the equivalent width limit begins to increase as the galaxies become fainter.
Galaxies are detected up to 1.25 magnitudes fainter if their line flux 
is stronger than $EW_o \sim 30$\angs. Some small fraction of the strong lined
objects at $m \approx 27$ are detectected, but the continuum is poorly defined.

\section{Properties of the [OII]-Selected Galaxy Population} \label{sec:results}

Considering the wide range of redshifts probed by an image of
this depth and the complexity of galaxy spectra, demonstration of an on-band flux
excess is not an immediately compelling argument for [OII] emission.  We use the
properties of the on-band selected galaxies and some basic information about the
field galaxy population to draw some conclusions about the contamination from
spectral features other than the [OII] doublet.  The population believed
to be [OII]-selected is then compared to the field population, and the 
distribution of star formation rates is discussed.

\subsection{The [OII] Line-Emission or Interloper Question}

\subsubsection{Emission-Line Interlopers}

Our on-band filter (W022) will pick up the  [OIII] $\lambda \lambda 4959, 5007$
doublet and the \Hb emission lines from galaxies near redshifts of 0.05 and
0.08, respectively. The volume probed for [OIII] and \Hb, however, is only
62\mpc3\ and 68\mpc3, respectively. Emission line surveys at $z \sim\ 0.1$ find 
roughly one galaxy per 25~Mpc$^{-3}$ (Salzer 1989), so we expect less than  5  
foreground galaxies to be selected.  The comoving volume from which 
redshifted [OII] emission is selected is 10 times larger. Since
the comoving density of emission-line galaxies also increases from z=0.1 to 
z=0.4, clearly few foreground galaxies will be found.

The redshift interval of our survey does probe a large volume at high-redshift,
and the issue becomes whether the galaxies are bright enough to detect in our survey.
The luminosity function of the $z \sim 3$ Lyman-break-selected population turns 
down at apparent magnitudes brighter than $\Re \sim 24.5$ (cf. Fig. 8 Steidel \et 1999a). 
These galaxies have $G - \Re$ colors ranging from 0 to 2, so most \lya\
detections will be from objects fainter than the completeness limit of our survey,
$G \approx m_{5129}(AB) \approx 24$. Figure~\ref{fig:on_l} shows the
lowest equivalent widths detected at $m_{5129} = 26.5$ and $m_{5129} = 29$
were 60\angs\ and 400\angs, respectively. 
About 20\% of the high-redshift galaxy population have rest-frame
\lya\ equivalent widths $\ge\ 20$\angs -- i.e. $EW_o$ exceeding $\sim 80$\angs\
(Steidel \et 1998; Steidel \et 1999b). Hence some $z \sim 3.31$ galaxies with strong 
\lya\ emission must be included in our catalog, but weaker lines like
CIII] $\lambda 1909$, or MgII~$\lambda 2798$ will be much less common.

The deepest \lya\ surveys find about 30 high-redshift
galaxies over volumes comparable to that probed by our on-band
(Cowie \& Hu 1998; Hu, Cowie, \& McMahon 1998).  
The sensitivity of our survey is lower, so we expect 
a smaller number of high equivalent-width  \lya\ emitters in our sample.
To a similar equivalent width limit, $EW_o = 100\angs$,  
our on-band-selected catalog contains only seven objects.  
These seven objects could all be [OII]-emitting galaxies 
in the cluster, however. Their rest-frame equivalent widths would be 
less than 285\angs. Galaxies have been found with [OII] 
equivalent widths as high as 264\angs\ (Sullivan \et 1999), and
widths of 100\angs\ are fairly common (e.g. Cowie \et 1996; 
Hammer \et 1997; Balogh \et 1997; Hogg \et 1998;
Tresse \et 1999; Gronwall 2000).  
The \lya\ candidates can be distinguished from the high-equivalent
width [OII] galaxies using the Lyman break (Lotz \& Martin 2001),
but a deep u-band image has not been obtained for \Ab851.
Our incompleteness function lacks the resolution to accurately estimate
the \lya\ contamination because the surface density of \lya\ emitters is rising 
steeply in the magnitude range where our sensitivity is dropping sharply.
We can conclude that the number of high-redshift \lya\ interlopers in
the on-band-selected sample is small regardless of the identity of 
the high-equivalent width lines.  The influence of \lya\ emitters
on the inferred [OII] equivalent width distribution is discussed further 
in \S~\ref{sec:ew}.

\subsubsection{Continuum Interlopers and Broadband Colors} \label{sec:colors}

The slope of a galaxy's continuum spectrum can produce a false equivalent width 
because the continuum is measured blueward of the on-band.  However,
 the {\it g} and {\it i} band fluxes place important constraints
on the degree of contamination.
To investigate the magnitude of the errors,  we
redshifted spectra of real galaxies (from our own collection, the
spectral atlas of Kennicutt (1992b), and the composite spectra of
Coleman, Wu, \& Weedman 1980) and ``observed'' them in our
four bands using the SYNPHOT package in IRAF.  Figure~\ref{fig:ew2}
illustrates the changes in apparent on-band excess as a function of redshift
for the Coleman \et (1980) representations of E/S0, Sbc, and Im galaxy spectra.
Weak [OII] lines in the Im and Sbc templates show up at z=0.407 as designed, but
the 4000\angs\ break in some foreground galaxies and the Lyman break in 
high-redshift galaxies also produce a detectable signal.  The broadband 
colors -- shown in the top panel of Figure~\ref{fig:ew2} -- provide a means of 
distinguishing interlopers from [OII]-emitters.

For example, early-type galaxies falling within several redshift intervals 
between z=0 
and z=1 produce an on-band flux excess exceeding $EW_o= 10$\angs\ but can be
recognized by their red \gi\ color which exceeds 2.0.  Our catalog contains
123 on-band-excess galaxies with $\gi\ > 2$, so the fraction of these which are actually 
interlopers does affect the statistics of the [OII]-selected population. The false 
equivalent width only exceeds $\sim 20$\angs\ when the steepest part of the break which 
is caused by metal lines in cool giants is redshifted across the filter pair. The
expected number of early-type galaxies within this redshift slice at  
$z \sim 0.3$ is of order  $\sim 12$  (Lilly \et 1995).
The number of interlopers with false $ew \sim 10$\angs\  is
much larger since galaxies from a much larger volume contribute.
We find 15 red galaxies with an on-band excess exceeding 30\angs. The number increases to 
48 red galaxies down to an equivalent width limit of 20\angs\ and to 116 down to
$EW_o = 10$\angs.  Either some of the red galaxies actually are
cluster galaxies with [OII] emission or the interloper equivalent width distribution
is offset by about 10\angs\ to higher equivalent width than our simulations
predict.  Given the wide variety of observed galaxy spectra, 
either interpretation is 
plausible.  We conclude that some of the red objects in the on-band
selected sample must be interlopers  but cannot reliably determine the fraction.

The Lyman discontinuity enters the off-band at z=4.56 and
exits the on-band at z=4.77  in \fig~\ref{fig:ew2}. The \gi\ color
of this population is quite red like the  4000\angs\ break interlopers, but
the magnitude of the signal is much larger.   The intrinsic
break is only about 70\angs\ for a pure O-star population but
becomes much larger if B stars dominate the continuum (Leitherer \et 1999).
Absorption by intergalactic gas along the sightline steepens the break
further and has been included in \fig~\ref{fig:ew2}  (Madau 1995). 
Since we find no high EW red galaxies, the Lyman discontinuity must not 
produce any interlopers in our sample.

\subsubsection{Spectroscopic Test of the [OII]-Selected Sample}

The validity of our technique can be checked by direct comparison to
the spectra recently published by  Dressler \et (1999).
They find 71 cluster members among spectra of 137 objects in their \Ab851
field but detect the [OII] emission line in only 19 of these members. 
Six of these 19 galaxies have equivalent widths below our sensitivity limit, and
the redshifts of four galaxies are outside the limits of our filter bandpass
(see  \S~\ref{sec:observations}).  Hence our sample should have nine galaxies
in common with the D99 sample. We identified eight of these galaxies as
 [OII] emitters.
The spectrum of the galaxy,  D99-\#52, that we missed is unusual in that
it shows a strong
 emission line shortward of [OII], and this line happened to fall
in our continuum band!\footnote{
 This line is near the wavelength of HeI 3554 from the galaxy, but 
it is much stronger than expected for HII galaxies.
The presence of this line in our continuum band explains why we measured
a weak off-band excess for this [OII] emitting galaxy.}
Figure~\ref{fig:ew_d99} compares our measurements of the [OII] equivalent
width to those of D99. The agreement is excellent except for one object,
D99-\#48, for which we detected only 25\% of the line flux. Examination of this
spectrum shows the [OII] line fell on the edge of the filter bandpass where
the transmission had dropped to $\sim 55\%$.  The root mean square deviation for the
other 7 points is 6\angs.

The false detection rate was examined for the 52 cluster galaxies with
no [OII] emission and the 66 field galaxies.  We measured an on-band
excess for 8 of these but flagged 6 of them as steep spectral breaks
rather than [OII] emission due to their red color (see \S~\ref{sec:colors}).
The other two had colors nearly as red
as the limit we used to make this distinction.  The detection algorithm
missed 3 galaxies that were located very close to bright
foreground stars. We measured no on-band excess for the other 104 objects.
This comparison, summarized in Table~2, demonstrates that the
selection function of our survey is well understood to $m_R(AB) \sim 22$.
Since the nature of systemic errors from interlopers changes with magnitude,
the method should be tested spectroscopically $m(AB) \sim 25$.

\subsection{Comparison to the [OII]-Selected Field Population}

Encouraged by the agreement with the published spectra, we consider
the galaxies selected in the on-band with $S/N \ge\ 3$ good candidates
for [OII] emission if their \gi\ color does not exceed 2.0.  This color
cut excludes 123 galaxies leaving 248 galaxies in the [OII]-selected sample.  
Direct comparisons to the  [OII]-selected population of
field galaxies near $z \sim 0.4$ (Hogg \et 1998 and Cowie \et 1997) 
are complicated by differences in selection criteria. However, 
internal comparisons to the off-band selected emission line galaxies
can be used as a cross-check on the comparison to the field population
of [OII]-emitting galaxies.

\subsubsection{The [OII]  Luminosity Function}

In Table~3 the galaxies from the [OII]-selected sample have been divided into four line 
flux intervals.  This method of counting adds faint galaxies with strong line emission to the
same bins as bright galaxies with weak line emission, so the affect of the survey selection function on the counts is not immediately apparent.  
We used the simulations described in  
\S~\ref{sec:incomplete} to measure the probability, $p_{ij}$, of detecting a galaxy with magnitude {\it i} 
and line flux {\it j}.  The number counts are then corrected by
the number of missed galaxies per detected galaxy, where this factor
is defined  by the sum,  $1/N\sum 1/p_{ij}$, 
 over all the galaxies in a bin containing N galaxies.
The correction factor for each flux bin is listed in column~5. 
The error estimate reflects  the dispersion  among several sets
of simulations.  We  find that  these observations provide a robust
measurement of the number of [OII]-emitting galaxies down to star-formation
rates  $\sim 0.3$\msunyr\ and  rest frame equivalent  widths
exceeding 8\angs. The counts are determined at the factor of two level
down to approximately 0.09\msunyr, but corrections for
galaxies fainter than the completeness limit of our survey 
strongly influence the number density in the  lowest  SFR bin.

The resulting distribution of star formation rates is shown in
Figure~\ref{fig:dsfr}.  The SFR scale includes a  mean correction for the 
filter transmission at the wavelength of the detected line -- i.e.
the measured  $T_{\lambda} F_{[OII]}$ was divided by  the average filter 
transmission $\bar{T_{\lambda}} = 0.78$ over the bandpass.
The exact redshift of a particular galaxy  within this interval 
$z = 0.407 \pm 0.0069$ has only a small effect on the its inferred luminosity.
While the narrow-band filter measurements appear to provide a robust
measurement of the reddened [OII] luminosities, differences in the
excitation among spiral and irregular galaxies create uncertainties 
in the conversion  to SFRs at the level of a factor of 2 to 3.  
Some galaxies will certainly be much more heavily obscured by dust than
the mean irregular galaxy, and their star formation rates would be
severely underestimated in the present analysis.
Until more robust  measures of total SFR can be obtained for these
galaxies, we can only assume that the  adjustments will shift the overall
scale of the SFRs and not their relative distribution.

To normalize the distribution in Figure~\ref{fig:dsfr} we
integrated the volume subtended by our field over the redshift
interval where the on-band filter transmits [OII] emission.  Dividing
the corrected number counts by this comoving survey volume, 1305~Mpc$^3$,
provides an estimate of the comoving density of galaxies at each SFR as
illustrated by the solid circles in \fig~\ref{fig:dsfr}.  For comparison,
the Hogg  \et (1998) [OII] luminosity function is shown to illustrate
the density of field galaxies at $z \sim 0.4$.  It agrees quite
well with the number density of [OII]-emitting galaxies  found by our 
off-band selected sample of emission-line galaxies (open circles). 
The comoving density of [OII]-selected galaxies is clearly a factor of 
three to four higher near \Ab 851 than it is in the field at similar
redshift.  This contrast is larger than the increase seen in the field
from $z \sim 0$ to $z \sim 0.4$ where the former is illustrated
by the \Ha derived SFR distribution of Tresse \et (1998) and Gallego \et (1995)
 in Figure~\ref{fig:dsfr}. Aside from normalization differences, the
shape of the SFR  distribution in the \Ab851 sample is very similar to that 
measured for the field galaxies.
Inspection of columns~3 and~4 in Table~3 reveals
that had we included the red galaxies with an on-band excess the shape of the
luminosity distribution would not be significantly affected although the overall
normalization would be up by about 50\%.  The overdensity is 70\% as large
when only the objects detected at $S/N \ge\ 4$ are included in the analysis,
and this difference provides at least some indication of how much the
magnitude of the overdensity might change when our spectroscopic follow-up is 
completed.

The peculiar motions of cluster galaxies clearly limit the accuracy with 
which the volume density of star-forming galaxies can be estimated. If the
velocity distribution of the cluster galaxies is described by a gaussian
distribution, then the random motions of 25\% of the cluster
galaxies will effectively remove them from our sample raising the true density
contrast with the field to a factor of five.  The chosen width of the
on-band filter still represents
a much longer distance along the sightline than the transverse size
of our field,  so our redshift slice may overestimate the volume in which
the cluster galaxies reside.
The field subtended at the cluster redshift is
 4.10~Mpc by 3.98~Mpc which is about 2 Abell radii across 
($R_A = 1.5 h^{-1} {\rm ~Mpc} \approx 2.1 {\rm ~Mpc} = 7\farcm1$).
Suppose, for purposes of illustration, that  all the [OII]-selected galaxies
lie within a volume whose depth along the sightline is also $2R_A$.  Then the 
comoving overdensity of star-forming galaxies increases to a factor of 
$\sim 42$.  The projected density of [OII]-selected galaxies at the outer 
edge of our field can be used to determine whether the overdensity of 
[OII]-selected galaxies is likely to be  confined to a region with dimensions 
of only $\sim 4$~Mpc however.

\subsubsection{Spatial Distribution} 

The top and bottom rows of \fig~\ref{fig:im_mo} show the spatial distribution 
of the on-band and off-band selected galaxies respectively.  The comoving 
volume probed by the off-band is 1.89 times larger than the on-band survey
volume, so the density contrast between the top and bottom panels is 
nearly twice as large as the contrast in galaxy counts. At clustercentric
radii of $\sim 2$~Mpc the cluster counts remain higher than the field counts,
so the scale of enhanced counts of [OII]-selected galaxies must be larger
than 4~Mpc.  We estimate that the virial radius of \Ab851 is $\sim 
4.6$~Mpc,\footnote{
	Estimates of the virial radius for \Ab 851 are uncertain
	due to the substructure in the X-ray emission (Schindler \& Wambssann 
	1996).  We
	modeled the halo as a singular isothermal sphere truncated at the 
	virial radius.
	Taking an isotropic velocity dispersion (1260\kms), we find a cluster 
	virial mass
	$M_V = 1.0 \times 10^{15} \msun\ R_V / R_A \sigma_{1000}^2$.   
	Following the
	collapse of a spherical perturbation, White \et (1993) find an 
	overdensity of
	$(\rho - \rho_0)/\rho = 178 \Omega_0^{-0.6}$ at virialization which 
	corresponds
	to a cluster mass $M_V = 6.9 \times 
	10^{14} h^{-1}\msun (R_V / R_A)^3 \Omega^{0.4}$.
	The only virial radius consistent with both these mass estimates is 
	$R_V \approx
	2.1 R_A \approx 4.6$~Mpc.}
so it seems reasonable that the field counts might not be recovered within
our field.

The radial gradient in the surface density of [OII]-selected galaxies is
not very steep, so the star-forming galaxies are not strongly clustered.
The projected density of star-forming galaxies in the inner half of our field 
is only $1.45 \pm 0.15$ times higher than in the outer half. The clustering
properties do not change significantly when only the sources detected
at $S/N \ge\ 4$ are included. The
radial surface density profile for a virialized population of 
galaxies would fall much more steeply with radius.  
The lack of a strong central concentration does not, however,
 demonstrate that this population is not virialized.
Star formation is believed to be strongly supressed in the galaxies that
pass through the core of the cluster, so galaxies would likely
become unrecognizable as members of this population in the cluster core.
A measurement of the redshift distribution of the [OII]-selected galaxies
will better determine  the dynamics of this population.

The spatial distribution of on-band selected galaxies does show more
structure than the off-band selected sample in \fig~\ref{fig:im_mo}.   
The galaxies with intermediate colors,  shown in the middle panel,
have concentrations near positions (1000,1000),  (850,1300) and (900,250)
which form a filament running roughly north -- south across the field.
The number of galaxies with blue colors is too low to determine if they are as 
clustered, but several  are located near the clumps at (1000,1000) and 
(900,250).  The apparent concentration of red galaxies along the 
vertical filament in Figure~\ref{fig:im_mo} disappears when selection 
is made at $S/N \ge\ 4$ instead of $S/N \ge\ 3$, so we find no
significant structure in the distribution of galaxies with red \gi\ color.

\subsubsection{Equivalent-Width Distribution} \label{sec:ew}

Abell 851 is an extraordinarily rich cluster, so the high density of 
[OII]-selected galaxies may only reflect the higher than average density 
of matter in this region.   The more interesting question is whether the 
star-formation histories of the [OII]-selected galaxies in \Ab851  are 
different from those of [OII]-selected galaxies drawn from the field.
The [OII] equivalent widths are the best measure we have at this time
of how quickly the star formation rate is changing in each galaxy.

The solid histogram in \fig~\ref{fig:ew} shows the distribution of $EW_o$ for 
the on-band-selected galaxies.  Most of the [OII]-selected galaxies in \Ab851 
have equivalent widths less than $EW_o = 80$\angs, but the seven galaxies add 
a flat high-equivalent width extension to the distribution.  The unusual
sharpness of this break led us to wonder whether an entirely different 
population of galaxies was being detected  at high equivalent width -- namely
the $z \sim3.31$ \lya\ emitters. If so, the same population should be seen in 
our off-band selected sample.  The open histogram in \fig~\ref{fig:ew}
shows the emission-line equivalent widths of the off-band selected galaxies.
Many more galaxies with equivalent-widths exceeding 100\angs\ were detected
in the off-band sample, and the distribution shows no discontinuity at 80\angs.
The emission-line galaxies are drawn from an off-band volume that is 
only two times larger than that of the cluster survey.
High-redshift galaxies are known to be strongly clustered
(Adelberger \et 1998), but it seems contrived to explain the high
number of large equivalent width galaxies in the off-band by invoking an 
overdensity  of galaxies at high-redshift.  

The straightforward explanation is that the high-equivalent population of 
[OII]-emitting galaxies has been removed from the cluster sample. 
Balogh \et (1998) found that cluster galaxies brighter than
$M_R \approx -18.3$ have lower [OII] equivalent widths than field 
galaxies of the same morphological type.  A direct comparison to
this work is difficult.  In \fig~\ref{fig:ew} our off-band selected 
sample cuts off in the $EW_o \sim 20 {\rm ~to~} 40$\angs\ interval
 instead of near $8 {\rm ~to~} 16$\angs\ like the 
on-band sample.  Weak lines become harder to detect when they fall 
in W021, the off-band, instead of W022. The continuum flux density is 
more uncertain when measured in the narrower band, and  this error is 
integrated over the broader bandpass when the continuum is subtracted.  
Detection at a fixed level of significance requires the line equivalent 
width to be a factor of two higher.  The Balogh \et sample of CNOC
field and cluster galaxies contains no high equivalent width galaxies.
We believe the difference mainly reflects our fainter magnitude limit.
If the seven  on-band galaxies with $EW_o > 100$\angs\ are at the
cluster redshift, their median absolute magnitude is only
$M_{\lambda 3645}(AB) = -14$.  Figure~8 of the Hogg \et (1998) field survey
shows the maximum [OII] equivalent widths increase with apparent magnitude.
The relative paucity of cluster galaxies with $EW_e > 60$\angs\ 
then reflects the recent suppression of star-formation in galaxies
fainter than the Balogh \et study reached.
We do attribute the $EW_o > 400$\angs\ off-band detections to
\lya\ interlopers, and their absence in the on-band sample to the smaller
volume.

\section{Star Formation History of [OII]-Selected Galaxy Population}

We have identified a population of star-forming galaxies
near the cluster \Ab 851 but commented little about the nature
of the underlying galaxies.  The types of star formation histories
represented are investigated here and used to discuss the nature
of the remnant populations in present-epoch clusters.

\subsection{Star Formation Timescales} \label{sec:birthrate}

 Figure~\ref{fig:ew_gi} shows the observed equivalent widths and \gi\ colors
of all the objects with an on-band excess.
The [OII] luminosities of galaxies are empirically related to the 
number of massive stars born over the last few Myr,  but the continuum 
emission near the line is dominated by stars with lifetimes of $\sim 2$~Gyr.
Their ratio, the line equivalent width, measures the current star formation
rate relative to the  average SFR over the previous few Gyrs. 
The \gi\ color also provides some age discrimination
because it spans the 4000\angs\ break of  redshifted cluster galaxies.
The strength of the break increases as stars of a few solar masses ascend 
the red giant branch and begin to dominate the continuum light, so the  
\gi\ color reddens about $\sim 500$~Myr after the stars form.  A large range
of both star formation timescales and population ages is required to explain 
the observed spread of equivalent width and color.

To describe the star-formation histories quantitatively,
the population synthesis code developed by Babul \& Ferguson (1996)
and  Madau \et (1996) was used to build continuum spectra of stellar
populations with different ages and star-formation histories. The
continua were reddened using an empirical description of the star -- dust 
geometry in starburst galaxies (e.g. Calzetti 1997; Calzetti \et 1996)
and redshifted to $z=0.4$. The g and i magnitudes were measured using 
the IRAF package SYNPHOT. The [OII] luminosity was estimated from the SFR 
parameter in each model using the relations described in \S~\ref{sec:sfr} and 
then attenuated for the assumed extinction, $A_B = 1.0$, using a Galactic 
extinction curve (Seaton 1979).  These extinction corrections reduce 
the [OII] equivalent width by a factor of 0.65 because the starlight 
suffers less extinction than the nebular lines for the starburst geometry.
The nebular continuum is not included in the current set of models.
Four different star-formation histories are illustrated 
in \fig~\ref{fig:ew_gi}.  The solid lines trace the aging population. 
Their absolute position would shift for different 
assumptions about the $L_{[OII]} / L_{\Ha}$ ratio or $A_B$, but their
overall shape serves to illustrate several important points.  

The solid squares along the top line in \fig~\ref{fig:ew_gi} follow a 
population with a constant star 
formation rate from an age of 4~Myr to 10~Gyr. The measured [OII] equivalent 
widths span a larger range than continuous star formation produces. The order 
of magnitude spread in their equivalent widths 
cannot be explained by extinction alone since 
some of the galaxies have very blue colors.  The most obvious explanation is 
that the star formation rate in many of the galaxies is declining with time.
The three evolutionary tracks below the constant SFR track illustrate the 
rapid decline in equivalent width with age when the star formation timescale 
is short. These stellar populations were created with an exponentially decaying 
SFR, $R = R_0 e^{-t/\tau}$, where $\tau =$ 10~Myr, 100~Myr, and 1~Gyr 
from left to right.  Decay  timescales $\tau \sim 10$~Myr to infinity (i.e.
continuous star formation) are needed to describe the range of colors and 
equivalent widths observed.

The equivalent width limit of our survey corresponds to a SFR at z=0.4 that is
a few percent of the average SFR over a period $\sim 6 \tau$. In other words,
the stars that dominate the light from the $\tau = 10$~Myr models
formed within the preceeding $ \sim 60$~Myr. It seems reasonable to call them
starburst galaxies.  In contrast, the longest decay timescales, $\tau \sim\ 
10^9$~yr,  are similar to the gas consumption timescales in local spiral galaxies 
(e.g. Kennicutt 1983; Kennicutt, Tamblyn, \& Congdon 1994).
Intermediate timescales may reflect the slow removal of gas by processes
specific to the cluster environment.  Hence more than one physical process is 
believed to be responsible for producing the wide range of star formation 
timescales.

 \subsection{Bursting Dwarf Galaxies?}

The 33 galaxies with the bluest \gi\ colors are highlighted by asterisks
in \fig~\ref{fig:ew_gi}.  They and a similar number of slightly redder, low 
equivalent width galaxies lie near the locus of $\tau = 10$~Myr models. 
The burst models redden slightly as their [OII] equivalent width decreases
along this track.  We will call these galaxies  fading starburst galaxies.  
Most of them are fainter than the cluster population that has been cataloged and
discussed previously (Dressler \et 1999; Smail \et 1997).

We can learn something about the stellar masses of the bursting galaxies
from their luminosities.  The best upper limit on their  K-band
luminosities -- $K \approx 18.0$ (Stanford, Eisenhardt, \& Dickinson 
1995)\footnote{
	Only 13 galaxies from the on-band excess sample were even detected
	in the infrared survey of Stanford, Eisenhardt, \& Dickinson (1995),
	and seven of these galaxies were flagged as probable foreground 
	interlopers.  One galaxy has no color measurement, and 
	the colors of the other 5 are between 1.6 and 2.0.}
-- is not faint enough to be a useful mass constraint. 
Our reddest flux measurement, sloan~{\it i},
reflects the luminosity emitted in the rest-frame V band. The mass
inferred from this luminosity depends on the star formation history adopted,
so we have tried to illustrate the range of masses that could reasonably
be assigned to each starburst galaxy in Figure~\ref{fig:mass}.  
Models with a constant star formation rate cross the solid heavy line at
the time when the total gas mass turned into stars is $10^9$\msun.  The 400~Myr 
old population with a SFR of 2.5\msunyr\ is bluer and brighter than a 
population of the same mass and age that formed quickly.  The fading sequence
for a burst that formed $10^9$ stars over $\tau = 10$~Myr, for example, is shown
by the heavy dashed line. After 10~Gyr, populations with a total stellar mass of
 $10^9$\msun lie along the dotted line connecting these two loci if their
star-formation timescales are intermediate to  the two extremes illustrated. 
  Congruent loci for $10^8$\msun, $10^{10}$\msun, and 
$10^{11}$\msun\ populations are outlined on the color -- magnitude diagram.
Comparison of the $10^9$\msun\ fading-burst track to the
$10^{10}$\msun\ constant SFR track clearly shows that galaxies at a given 
magnitude and color will be assigned a lower stellar mass if a fading burst 
model rather than a constant SFR model is invoked.  The interesting point
is that the upper limit on the masses of the bluest starburst galaxies
is only $10^9$\msun, and the best estimate of the stellar mass involved
in the burst is closer to $10^8$\msun\ models.  
These galaxies could be dwarf galaxies.

The dwarf galaxy candidates are smaller than a typical [OII]-selected
galaxy.  At the detection threshold, the isophotal area of a galaxy 
with $\gi\ < 1.0$ is typically a factor of two smaller than one with 
$ 1.0 < \gi\ < 2.0$. This trend reflects a tighter underlying
correlation between size and magnitude. All of the faint galaxies
are small, but few of the brighter [OII]-selected galaxies have an
extremely blue \gi\ color. We detect cluster galaxies in their rest-frame
U band, so {\it size} here is understood to imply {\it size of the
star-forming region}.  The nearly linear correlation between size
and flux implies the star-forming regions in the dwarfs have 
the same central surface brightness but smaller scale length than the
brighter galaxies. These bursts of star-formation activity may
briefly outshine an older stellar population without contributing much to the 
galaxy's total stellar mass.  Gas-phase metallicity measurements will be 
the best way to confirm or refute the dwarf-identity of these galaxies.
We note that the oxygen abundances recently measured for a few
compact galaxies in Cl~0024+16 implied higher stellar masses than
either the luminosities or line widths of those galaxies revealed
(cf. Kobulnicky \et 1999; Koo \et 1997).

\subsection{Implications for the Formation of the Dwarf Galaxy Population
in Clusters}

If the low [OII] equivalent widths of the cluster galaxies relative to 
galaxies in the field is taken as evidence that the SFR's in the cluster 
galaxies are declining faster than normal, then  some environmental process
must be removing gas from them.  The physical cause of the gas removal is
still unclear, but galaxy harrassment appears to be consistent with the lack of
a strong clustercentric gradient in galaxy properties and a broad range of decay 
timescales for the SFR (Moore \et 1996).  

Most of the short bursts appear to have taken place in dwarf galaxies.  
The luminosity of their remnants depends quite strongly on the star 
formation history before and after the luminous burst. However, the maximum 
fading rate of the bursting galaxies and their minimum remnant luminosities
follow directly from the SFR and birthrate parameter measured at z=0.4. 
Such fading starbursts have been discussed frequently in the literature as
a means of lowering the luminosity of prominent high-redshift galaxies
below detection limits by the current epoch (Phillipps \& Driver 1995;
Ferguson \& Babul 1998).   The R-band absolute magnitudes of remnants produced 
by the star formation histories defined in \S~\ref{sec:birthrate} 
are compared in  Table~4.  A bursting dwarf galaxy with a SFR of $0.1$\msunyr\
at $z = 0.4$ has faded to $M_R = -10.13$ to  -9.6 by the present epoch, and this range
represents metallicities from 0.02\zsun and 1.0\zsun. In contrast, if small 
bursts propagate over the galaxy such that the 
galactic SFR is nearly constant at $0.1$\msunyr, then the remnant's 
luminosity increases to $M_R = -16.55$ to -17.73 for the same metallicity interval.
We cannot directly distinguish between these two scenarios because
it is not possible to constrain the number of bursts in an individual object 
with the present data.

In spite of the limited information, it is interesting to compare models for
the luminosity distribution of the remnants to the dwarf population in nearby 
galaxy clusters. Since \Ab 851 is a richer cluster than  Virgo, a plausible model 
should produce at least as many dwarfs of a given type as are found in the 
Virgo Cluster Catalog, VCC (Sandage \et 1985).  The VCC lists
34 blue compact dwarfs, over 100 dwarf irregular galaxies, and about 200 spiral
galaxies.  These numbers are not very different from the number of 
blue, compact galaxies (33) and moderately blue galaxies (160) that we found
in \Ab851.  However, the star formation rates in the [OII]-selected galaxies 
cannot remain constant at their $z=0.4$ rate.  Most of the [OII]-selected galaxies,
i.e. those with moderately blue \gi\ color, would become as luminous as the
spiral galaxy population in Virgo.  While this constant star formation rate scenario
does not produce too many spiral galaxies, it does predict $z=0$ star formation rates
that are higher than those of typical Virgo cluster spiral galaxies.  Also, 
nearly half of the galaxies in Table~1 would be brighter than the 
VCC dwarf galaxies, so the progenitors of the Virgo dwarf population would remain unidentified.
In constrast, the declining star formation rates in \Ab851 galaxies
suggested by our analysis of their color and equivalent width distribution  
produces many more dwarf galaxy remnants.  We would only classify a galaxy as a burst 
during a 60~Myr period, so the number of remnants from galaxies that undergo a starburst over the 
$\sim 7$~Gyr period from redshift zero to $z = 1$ could be two orders of magnitude larger than 
number directly detected.

For purposes of illustration, we assume the main star-formation episode has been 
detected in the [OII]-selected galaxies.   Which remnants would then have luminosities 
similar to the dE population in clusters at the present epoch? The top panel and middle panel of
Figure~\ref{fig:fade_nm} illustrate the remnants of the population that was
[OII]-selected at z=0.4 with $\gi < 2.0$.  Many galaxies assigned short, $\tau = 10$~Myr,
SFR decay timescales would have faded below the detection limit of the VCC, and most
galaxies brighter than $M_R = -13$ would be descendents of the population with 
a slower supression of the SFR. These galaxies had redder \gi\ colors at z=0.4 similar 
to spiral galaxies, so the implication is that many of the dE-progenitors underwent
multiple generations of star-formation.  These more luminous progenitors may not create
a metallicity problem for the remnants.  The J-K colors of some Virgo dE's indicate
fairly high metallicities (Bothun \et 1985), and
a metallicty range from 0.02\zsun\ to 1.0\zsun\ would produce a range of $b-r$
colors consistent with the color spread observed among the dE 
galaxies in the Coma cluster (Secker \& Harris 1996).

The problem with forming stars late ($z < 1.0$) in dE progenitors seems to be
producing enough remnants. The bottom panel of Figure~\ref{fig:fade_nm} 
illustrates the situation.  The dE population was assumed to form in the cluster
between $z = 1.0$ and $z= 0.4$.  The histograms for the three star formation
timescales shown in the middle panel were then scaled by their respective duty cycle 
corrections.  Galaxies  with star formation timescales $\sim 10$~Myr produce more remnants
than those with longer star formation timescales when the remnant
population is boosted by the duty cycle correction.  However,
the total number of gas-depleted remnants in each magnitude bin, dotted line, is still less than
the estimated number of Virgo dE's, solid line.\footnote{
	We  have assumed a distance to the Virgo  Cluster of 17.0~Mpc and a 
	mean B-R color of 1.32 since the VCC lists only B magnitudes.}  
Over the range from $-18 < M_R < -13$, only 277 remnants are produced whereas the VCC 
lists about 1000 dE galaxies in this magnitude range. 
To resolve the numbers problem the progenitors of dE galaxies would need to form stars over
a longer period from $z \sim 2$  to $z \approx 0$ (yielding over 800 remnants in the above
magnitude range).  Perhaps this period reflects the timescale over which the cluster
environment has been able to supress the star formation activity of infalling galaxies.

In summary, our impression is that \Ab851 does not contain enough bursting 
dwarf galaxies to form a large dE galaxy population by the present epoch.  Either
many dE progenitors are not dwarf galaxies or one of two scenarios holds.  First,
it cannot be ruled out that the dwarf galaxies in \Ab851 went through a bursting phase
at an earlier epoch, prior to $z = 0.4$.  This solution is unsatisfying unless
it can be connected to some dynamical aspect of cluster assembly.  Second, we 
assumed the dwarf galaxies experienced only a single burst of star formation 
activity.  In a more realistic model where dwarf galaxies burst multiple times,
the remnants will be brighter.  Although their duty cycle is also larger in
this scenario, the net number of detectable remnants can be increased over the
single-burst scenario.  It is likely that our [OII]-selected population includes
more than one type of dE progenitor -- the dwarf galaxies which have undergone these
multiple episodes of star formation and some brighter, spiral galaxies which
will be partially disrupted by tidal interactions.  Despite the large number of 
dwarf and spiral galaxies with declining star formation rates in \Ab851, an extended 
period  $\sim 8-9$~Gyr seems to be required to build up the dE population.

\section{Summary}

We presented a deep, emission-line-sensitive image of the galaxy cluster \Ab851.
Galaxies with strong [OII]$\lambda\lambda 3726,29$ emission 
($EW_o \sgreat\ 11$\angs) were detected to $m_{5129}(AB) = 24.0$.
Foreground galaxies with steep  breaks in the spectral continuum near 
4000\angs\ contaminate the sample selected by on-band excess flux.
We showed, however,
 that such interlopers will have very red \gi\ colors and excluded 
123 on-band-selected galaxies with red colors
 from the sample of 248 probable  [OII] emitters.
 We argued that high-redshift galaxies with extremely strong
\lya\ emission must contaminate the sample, but the differences
between the equivalent width distributions of the on-band-selected
and off-band-selected samples lead us to believe most lines
with $EW_o < 400$\angs\ (i.e. $EW_e < 285$\angs) are  [OII] detections.

Many of the [OII]-selected galaxies are too faint to have previously 
measured redshifts.  We constrain their redshifts to the interval 
$0.399 < z < 0.415$, which includes cluster members with peculiar
velocities from $-1474$\kms to $1450$\kms and field galaxies within 
several tens of Mpc of the cluster along our sightline.
Their co-moving density is  at least several times higher than that 
of [OII]-selected field galaxies at similar redshifts 
(Cowie \et 1997; Hogg \et 1998).  Hence the [OII]-selected population 
appears to be associated with the overdensity of  galaxies that defines 
the cluster even though their projected spatial distribution is less
centrally concentrated than the general cluster population.  We suspect
the extra star-forming galaxies are associated with infalling
galaxies over scales $\sgreat\ 2$~Mpc, the radius of our field of view.

The star-formation episodes within individual galaxies are not as 
strong on-average as the bursts seen in the field population. The 
cluster galaxies have lower equivalent widths than [OII]-selected 
field galaxies. The range of equivalent widths and \gi\ colors among 
cluster galaxies is significantly larger than differences in intrinsic 
reddening and/or excitation (i.e. the scatter in the [OII]-\Ha relation) can 
easily explain.  We believe galaxies with very different types of star-formation
histories are represented in the [OII]-selected population.
We used population synthesis models to demonstrate that the star formation
timescales in two-thirds of the [OII]-selected sample are of order a few
times $10^8$ to $10^9$~yr.  These galaxies have moderately blue colors 
similar to normal spiral galaxies,  and the high end of this range is
consistent with the gas consumption timescales in field spirals (Kennicutt
\et 1994; Kennicutt 1983).  The intermediate timescales may reflect
the slow removal of gas by the cluster, but we regard these timescales 
tentatively until confirmed using other spectral diagnostics less sensitive
to reddening.  In contrast, the star-formation timescale of at least $13\%$, 
and possibly of as many as one third, of the galaxies is distinctly shorter.
These galaxies are very blue ($\gi\ < 1.0$), and we believe they
are starburst galaxies. 

While the luminosity of these starburst galaxies is clearly dominated by a 
young population of stars, estimates of their total stellar mass are highly 
uncertain because the star formation history more than a few Gyr prior to the 
$z \sim 0.4$ epoch is not well constrained by the available data.  Nonetheless
the mean i-band magnitude of the starburst population is $\sim 2$ magnitudes 
fainter than that of the  spiral-like population, and the absolute luminosities 
of the starburst galaxies range from $M(AB) \approx -18$ to $-14$ (rest-frame U
band). 
Their stellar masses could be as low as $ 10^8$ to $10^9$\msun, so we 
we suggest they are bursts of star formation in dwarf galaxies. 
A preliminary discussion of the sizes of these galaxies showed they were 
among the most compact objects selected as [OII]-emitting galaxies.  
Spectroscopic observations will yield metal abundances and absorption-line 
strengths for this population which will reveal their true nature.

We find there simply are not enough of these small starbursts to form all
of the dE's seen in nearby clusters via a wind-driven transformation of
gas-rich dwarfs into dE galaxies. The bursting  dwarfs do form an important 
subset of a larger population of [OII]-selected galaxies which must 
signficantly decrease their SFR's by the current epoch however. Otherwise far
more bright spirals and irregulars would be found in clusters today. 
We  have caught these [OII]-selected galaxies at a stage when their SFR 
is declining faster than it would in the field. While their projected 
spatial distribution is not as centrally concentrated as the nucleated dE
population in the Virgo Cluster, the density contrast across our field
of view is very similar to that measured for the bright dE population
in the Virgo cluster over comparable length scales (Binggeli \et 1987;
Ferguson \& Sandage 1989). Our analysis certainly suggests then that 
dwarf elliptical galaxies were forming in clusters at $z \approx 0.4$.

\acknowledgements{We thank Ed Carder of KPNO for measuring the filter
transmission curves and simulating their response in the f/3.14 beam prior to 
our run. We express our sincere thanks to Daniela Calzetti for illuminating
discussion about the impact of dust on the [OII] equivalent width.  We
would to thank the Morphs group 
(Dressler, Oemler, Couch, Ellis, Poggianti, Barger, Butcher,
Sharples, Smail) for creating an easily accessible WWW
interface to their \Ab851\ spectra. This work was supported by NASA
through Hubble Fellowship grant HF-01083.01-96A awarded by the Space
Telescope Science Institute, which is operated by
the Association of Universities for Research in Astronomy, Inc., for NASA.
}


\begin{deluxetable}{rrrrrrrrr}
\tablecaption{Emission-Line Candidates with Blue Colors}
\tablehead{
\colhead{ID\#} &
\colhead{X}  &	
\colhead{Y}  & 
\colhead{RA }  & 
\colhead{DEC }  & 
\colhead{T$_{\lambda}$ F} &
\colhead{F$_{\lambda}$}	& 
\colhead{g} & 
\colhead{g-i} \\
\colhead{}   &
\colhead{ (pix)}	&	
\colhead{  (pix)}	& 
\colhead{  (j2000)}	& 
\colhead{  (j2000)}	& 
\colhead{ (ergs~s$^{-1}$~cm$^{-2}$)} &  
\colhead{(ergs~s$^{-1}$} &
\colhead{(mag)} &
\colhead{(mag)} \\
\colhead{}   &
\colhead{}  &	
\colhead{}  &	
\colhead{}  &	
\colhead{}  	& 
\colhead{}   &  
\colhead{cm$^{-2}\AA^{-1}$)} &  
\colhead{}	 & 
\colhead{}
}
\startdata
     2413 &         939.58 &         832.64 & 09:42:59.528 & +46:57:27.68  &  $ 9.53 \pm  0.58\times 10^{-17}$& $   1.41\times 10^{-18 }$ &      23.71 $\pm$ 0.23&        0.64  $\pm$     0.29  \\
     5706 &         1115.7 &        1258.42 & 09:42:52.381 & +47:00:24.01  &  $ 1.180 \pm 0.073\times 10^{-16}$&$    1.99\times 10^{-18}$ &       23.43 $\pm$0.22&        0.85  $\pm$     0.28  \\
     4230 &         764.56 &         1423.8 & 09:43:06.763 & +47:01:32.23  &  $ 1.258 \pm 0.078\times 10^{-16}$&$    2.07\times 10^{-18}$ &       23.45 $\pm$0.23&        0.77  $\pm$     0.29  \\
      368 &        1024.44 &         130.81 & 09:42:56.013 & +46:52:40.81  &  $ 1.38 \pm  0.11\times 10^{-16}$& $   3.54\times 10^{-18 }$ &      22.93 $\pm$ 0.22&        0.68  $\pm$     0.28  \\
      756 &         525.94 &         265.14 & 09:43:16.228 & +46:53:34.91  &  $ 1.81 \pm  0.15\times 10^{-16}$& $   4.98\times 10^{-18 }$ &      22.59 $\pm$ 0.22&     0   $\pm$    0.29  \\
      281 &        1491.63 &          96.78 & 09:42:37.113 & +46:52:28.03  &  $ 9.01 \pm  0.75\times 10^{-17}$& $   2.18\times 10^{-18 }$ &      23.4 $\pm$ 0.22&         0.74  $\pm$     0.29  \\
     5740 &          21.04 &        1269.62 & 09:43:37.124 & +47:00:27.58  &  $ 1.19 \pm  0.10\times 10^{-16}$& $   3.23\times 10^{-18 }$ &      23.05 $\pm$ 0.22&        0.87  $\pm$     0.28  \\
      773 &         930.19 &         273.34 & 09:42:59.844 & +46:53:38.57  &  $ 6.01 \pm  0.64\times 10^{-17}$& $   1.55\times 10^{-18 }$ &      23.88 $\pm$ 0.23&        0.67  $\pm$     0.30  \\
     3392 &        1881.27 &        1124.57 & 09:42:21.137 & +46:59:30.07  &  $ 3.89 \pm 0.47\times 10^{-17 }$& $  1.02\times 10^{-18  }$ &     24.65 $\pm$ 0.25 &       0.73   $\pm$    0.32  \\
     3157 &        1595.69 &        1056.04 & 09:42:32.786 & +46:59:01.07  &  $ 8.1 \pm  1.0\times 10^{-17  }$& $ 3.43\times 10^{-18   }$ &    22.92 $\pm$ 0.22 &         0.51  $\pm$     0.28  \\
      522 &         916.12 &          192.5 & 09:43:00.405 & +46:53:05.69  &  $ 5.11 \pm  0.65\times 10^{-17}$& $   1.71\times 10^{-18 }$ &      23.85 $\pm$ 0.23&         0.89  $\pm$     0.29  \\
     2303 &         955.22 &         803.64 & 09:42:58.887 & +46:57:15.76  &  $ 1.76 \pm 0.29\times 10^{-17 }$& $  4.5\times 10^{-19  }$ &     25.32 $\pm$  0.26 &       0.48 $\pm$      0.36  \\
\tablebreak
     5269 &        1787.81 &        1757.03 & 09:42:24.896 & +47:03:52.52  &  $ 2.13 \pm  0.397\times 10^{-17}$&$    8.1\times 10^{-19}$ &       25.34 $\pm$  0.28&      0.70 $\pm$       0.39  \\
     3954 &         358.89 &        1344.21 & 09:43:23.341 & +47:00:58.74  &  $ 1.38 \pm 0.25\times 10^{-17 }$& $  3.2\times 10^{-19  }$ &     25.99 $\pm$ 0.32 &       0.86  $\pm$     0.43  \\
      824 &         885.24 &          301.2 & 09:43:01.670 & +46:53:49.86  &  $ 1.86 \pm 0.37\times 10^{-17 }$& $  6.8\times 10^{-19  }$ &     24.77 $\pm$ 0.24 &        0.66 $\pm$      0.32  \\
      459 &        1931.46 &         167.47 & 09:42:19.316 & +46:52:58.12  &  $ 4.51 \pm  0.93\times 10^{-17}$& $   3.18\times 10^{-18 }$ &      22.78 $\pm$ 0.22&        0.72  $\pm$     0.28  \\
     4829 &         938.56 &        1645.91 & 09:42:59.673 & +47:03:04.92  &  $ 2.68 \pm  0.62\times 10^{-17}$& $   1.88\times 10^{-18 }$ &      23.58 $\pm$ 0.22&      -0.04  $\pm$     0.30  \\
      889 &         909.21 &           328. & 09:43:00.702 & +46:54:00.81  &  $ 9.3 \pm 2.3\times 10^{-18   }$& $2.6\times 10^{-19    }$ &   26.22 $\pm$ 0.34  &        0.83 $\pm$       0.47  \\
     2711 &         941.76 &         923.98 & 09:42:59.450 & +46:58:05.35  &  $ 4.31 \pm 1.07\times 10^{-17 }$& $  3.87\times 10^{-18  }$ &     22.85 $\pm$ 0.22 &        0.25  $\pm$     0.28  \\
       89 &        1811.27 &          25.84 & 09:42:24.211 & +46:52:00.38  &  $ 1.93 \pm  0.48\times 10^{-17}$& $   1.17\times 10^{-18 }$ &      24.89 $\pm$ 0.26&        0.42  $\pm$     0.37  \\
     5066 &        1953.46 &        1736.64 & 09:42:18.121 & +47:03:44.31  &  $ 9.39 \pm 2.42\times 10^{-17 }$& $  9.48\times 10^{-18  }$ &     22.00 $\pm$  0.22 &       0.85   $\pm$    0.28  \\
     2334 &         351.07 &         809.65 & 09:43:23.497 & +46:57:17.71  &  $ 1.79 \pm  0.47\times 10^{-17}$& $   1.09\times 10^{-18 }$ &      24.43 $\pm$0.24&        0.26 $\pm$      0.33  \\
     2524 &         663.91 &         875.69 & 09:43:10.769 & +46:57:45.12  &  $ 1.33 \pm  0.35\times 10^{-17}$& $   6.3\times 10^{-19 }$ &      24.72 $\pm$ 0.24&        0.63 $\pm$      0.32  \\
     3815 &        1055.11 &        1256.43 & 09:42:54.858 & +47:00:23.10  &  $ 8.7 \pm 2.3\times 10^{-18   }$& $3.0\times 10^{-19    }$ &   25.72 $\pm$ 0.28   &       0.61 $\pm$       0.38  \\
\tablebreak
      737 &        1134.09 &         272.67 & 09:42:51.580 & +46:53:38.61  &  $ 1.04 \pm 0.28\times 10^{-17 }$& $  4.0\times 10^{-19  }$ &     25.63 $\pm$ 0.29 &       0.81   $\pm$    0.38  \\
      508 &         944.84 &          188.9 & 09:42:59.241 & +46:53:04.27  &  $ 1.54 \pm 0.42\times 10^{-17 }$& $  8.1\times 10^{-19  }$ &     24.90  $\pm$ 0.26 &       0.80  $\pm$     0.34  \\
     1118 &         701.44 &         405.99 & 09:43:09.144 & +46:54:32.41  &  $ 9.4 \pm 2.7\times 10^{-18   }$& $3.8\times 10^{-19    }$ &   25.69 $\pm$ 0.29   &       0.80     $\pm$  0.39  \\
     3733 &        1483.21 &        1227.03 & 09:42:37.367 & +47:00:11.63  &  $ 1.60 \pm  0.46\times 10^{-17}$& $   1.09\times 10^{-18 }$ &      24.39 $\pm$ 0.24&        0.88 $\pm$      0.30  \\
     4319 &         168.76 &        1449.74 & 09:43:31.150 & +47:01:42.41  &  $ 1.65 \pm 0.47\times 10^{-17 }$& $  1.11\times 10^{-18  }$ &     24.27 $\pm$ 0.24 &           0.25 $\pm$      0.32  \\
     1889 &        1269.36 &         671.82 & 09:42:46.092 & +46:56:22.06  &  $ 7.0 \pm 2.0\times 10^{-18   }$& $1.7\times 10^{-19    }$ &   26.0 $\pm$ 0.30  &       0.34 $\pm$  0.46 \\
     2987 &          930.6 &         992.86 & 09:42:59.913 & +46:58:33.78  &  $ 1.96 \pm  0.58\times 10^{-17}$& $   1.33\times 10^{-18 }$ &      24.02 $\pm$ 0.23&        0.83  $\pm$     0.30  \\
     1337 &         160.92 &         473.73 & 09:43:31.105 & +46:54:59.95  &  $ 1.82 \pm  0.55\times 10^{-17}$& $   1.43\times 10^{-18 }$ &      24.30 $\pm$ 0.24&       -0.46  $\pm$     0.41  \\
     2196 &         551.66 &         765.95 & 09:43:15.316 & +46:56:59.84  &  $ 1.95 \pm  0.61\times 10^{-17}$& $   1.80\times 10^{-18 }$ &      23.42 $\pm$ 0.22&        0.57  $\pm$     0.29  \\
\enddata
\end{deluxetable}



\begin{table}
\caption{Comparison to Spectroscopic Subsample\tablenotemark{a}}
\begin{tabular}{lll}
\hline
$N_{gal}$ 	& $N_{gal}$		& Description \\
D99	&	this paper  & \\
\hline
\hline
71		&		& Identified as Cluster Members\tablenotemark{a} \\

19		&		& Detected [OII] in Spectrum\tablenotemark{a} \\
		& 8		& Identified as [OII] Candidates\tablenotemark{b}\\
		& 4		& Redshifted beyond filter bandpass.\tablenotemark{b}\\
		& 6		& Equivalent width less than  (12\AA).\tablenotemark{b}\\
		& 1		& Unusual emission-line spectrum.\tablenotemark{c}\\  
52		&		& No [OII] Detected in Spectrum\tablenotemark{a} \\
		& 45		& No on-band excess detected \\
		& 5		& On-band excess significant. Too red ($g-i > 2.0$) for EL \\
		&		& ~~~galaxy sample (4); $g-i = 1.79 \pm 0.28$ (1). \\
		& 2		& Detected but not extracted by SE. \\
66		&		& Classified as field galaxies.\tablenotemark{a} \\
		& 59		& No on-band excess\tablenotemark{b} \\
		& 3		& Significant on-band excess.  Too red for EL sample (2)\\
		&		& ~~~and one with $g-i = 1.71 \pm 0.28$. \\
		& 3		& Detected but not extracted by SE. \\
		& 1		& Miss-classified as a star. \\
\hline
\end{tabular}
\tablenotetext{a}{Dressler et al. 1999.}
\tablenotetext{b}{This paper.}
\tablenotetext{c}{The [OII] equivalent width should be easily detected, yet
we measured a negative line flux.  See text for details.
}
\end{table}



\begin{table}
\caption{Distribution of [OII] Emission-Line Fluxes}
\begin{tabular}{llllll}
\hline
$\log (T_{\lambda} F_l)$\tablenotemark{a} &
	$\dot{M}$\tablenotemark{b} & 
		$N$\tablenotemark{c} & 	  		
			$N_{oii}$\tablenotemark{d} & 	  		
				$C_f$ \tablenotemark{e} & 	  		
					$\eta$\tablenotemark{f}\\
\hline
\hline
-15.25	& 5.30	& 17	& 9  	& $1.02 \pm 0.02$ & $6.7 (\pm 2.2) \times 10^{-3}$	\\
-15.75	& 2.00  & 62 	& 42	& $1.05 \pm 0.03$ & $3.2 (\pm 0.5) \times 10^{-2}$	\\
-16.25  & 0.53	& 107   & 71	& $1.10 \pm 0.03$ & $5.7 (\pm 0.7) \times 10^{-2}$	\\
-16.75  & 0.17 & 147   & 91	& $2.63 \pm 0.9$  & $0.17 \pm 0.06$	\\
\hline
\end{tabular}
\tablenotetext{a}{Measured line flux in ~ergs~s$^{-1}$~cm$^{-2}$.
These values must be divided by the filter transmission at the wavelength
of the line.
}
\tablenotetext{b}{Estimated star formation rate in $~{\rm M_\odot}$~yr$^{-1}$.
See text for details.}
\tablenotetext{c}{Total number of $3\sigma$ detections.}
\tablenotetext{d}{Number of objects in column~3 which are likely to be detections
of [OII] emission.   The 123
galaxies with red g-i colors and the 7 galaxies with equivalent widths 
exceeding 100\AA\ are excluded.}
\tablenotetext{e}{Incompleteness correction.  Each galaxy was assigned a detection
probability, $f_i$, interpolated from a grid of Monte Carlo simulations.  The average
correction for each bin, $1/N \sum 1/f_i$, is given.  The error bars represent the uncertainty 
in this interpolation or the Poisson noise, whichever is larger.}
\tablenotetext{f}{Number of emission-line galaxies per Mpc$^{-3}$/0.5~mag
(\h0\ = 70 km s$^{-1}$ Mpc$^{-1}$, $\Omega_M = 0.2$, and $\Omega_{\lambda} = 0$).
}
\end{table}



\begin{table}
\caption{Population Synthesis Models\tablenotemark{a}}
\begin{tabular}{lrllrlll}
\hline
$\tau$\tablenotemark{b}  & $t_o(0.4)$	& g-i	& $EW_o$ & $t_o(0)$ &0.02\zsun\ & 0.2\zsun\	& 
1.0\zsun	\\
Myr	& Myr	& mag	& $\AA$   &  Gyr            & b-r, $M_R$ &
b-r, $M_R$ & b-r, $M_R$ \\
\hline
\hline
 10	& 20	& 0.48	& 93 & 4  & 1.05, -10.13& 1.29, -9.88& 1.59, -9.6	\\
 10	& 40	& 0.69	& 30 & 4  & 1.05, -10.13& 1.29, -9.88& 1.59, -9.6	\\
 10 	& 60	& 0.91	& 7  & 4  & 1.05, -10.13& 1.29, -9.88& 1.59, -9.6	\\
 100	& 200	& 1.02	& 65 & 4  & 1.06, -12.64& 1.29, -12.39&	1.60, -12.1	\\
 100	& 400	& 1.61	& 25 & 4  & 1.06, -12.64& 1.29, -12.39&	1.60, -12.1	\\
 100	& 600	& 2.00	& 7  & 4  & 0.95, -15.49& 1.29, -12.39&	1.60, -12.1	\\
 1000	& 2000	& 1.71	& 53 & 4  & 0.95, -15.49& 1.11, -15.26&	1.28, -14.95	\\
 1000	& 4000	& 2.35	& 26 & 8  & 0.95, -15.49& 1.35, -19.47&	1.62, -19.05	\\
 1000	& 6000	& 2.82	& 9  & 8  & 1.13, -19.73& 1.35, -19.47&	1.62, -19.05	\\
 $\inf$ & $10^4$ & 1.64	& 68 &14  &0.78, -16.55&0.88, -18.00& 0.93, -17.73	\\
\hline
\end{tabular}
\tablenotetext{a}{
All models include an intrinsic reddening of $A_B = 1.0$~magnitude
and are normalized to a star formation rate of 0.1{\mbox {$~{\rm M_\odot}$~yr$^{-1}$}}.}

\tablenotetext{b}{Col. 1 -- The decay timescale for the star formation rate where
the rate is $R = R_0 e^{-t/\tau}$.
Col. 2 -- The age of the stellar population at redshift 0.4.
Col. 3, 4 -- The color and [OII] emission line equivalent width measured in the
observer's frame.
Col. 5 -- The age of the stellar population at redshift zero.
Col. 6 - 11 -- The color and absolute magnitude of the remnant for three values
of the metallicity.}
\end{table}

\clearpage

\begin{figure} 
	\caption{Filter transmission in the f/3.14 beam.  The bandpass of
filter W022, our {\it on band},  includes [OII] $\lambda \lambda 3726,29$ 
emission at the 
cluster redshift.  We estimate the continuum level from the flux measured in
W021, our {\it off band}.}
	\label{fig:filters} \end{figure}

\begin{figure} 
	\caption{The central region of the co-added narrow band images.
The lower left corner of this panel has coordinates (659,659) in the
system of image coordinates used in Table~1. Tickmarks are shown at 20 pixel intervals.
Galaxies with an  on-band excess at the $S/N \ge\ 3$ level are marked according to their
\gi\ color -- $\gi\ < 1.0$ (open squares), $1.0 < \gi\ < 2.0$ (open ciricles),
$\gi\ > 2.0$ (X's), and poorly measured color (open triangles).  Galaxies with
a flux excess in the off-band are only indicated if their color falls in 
the bluest bin (open squares with an X).  The other 8 panels of the field can
be viewed at http://astro.caltech.edu/~clm/dwoii.html.  
}
	\label{fig:sum} \end{figure}

\begin{figure} 
	\caption{Star -- galaxy separation.  Filled circles denote objects
classified as stars based on their area (at a specified surface brightness)
and flux.}
	\label{fig:stars} \end{figure}

\begin{figure} 
	\caption{Signal-to-noise ratio of the on-band excess vs. continuum flux.
The sensitivity of our survey to emission lines in the on band is shown by
the solid lines for four galaxy models:   (a) emission-line equivalent width of 
300 \AA\ and area of nine pixels, (b) emission-line equivalent width of 60\AA\ and
area of nine pixels, (c) emission-line equivalent width of 60\AA\ and area of 200
pixels, and (d) emission-line equivalent width of 11\AA\ and area 200 pixels.
A selection cut at $S/N = 4$ is equivalent to an equivalent width cut at 11\AA\ for 
small galaxies with $\log F_{\lambda} = -18$ which is $m_{5129}(AB) = 24.0$. The
dotted line at $S/N = 3$ illustrates the selection criterion used for Table~1.
}
	\label{fig:fsig} \end{figure}

\begin{figure} \caption{On-band flux excess vs continuum flux density
at $\lambda 5129$. The diagonal lines show locii of constant equivalent width 
at 100, 60,  and 20 \angs, from top to bottom, for reference.  The signal-to-noise
level of the limiting on-band flux excess shown, $S/N \ge\ 3$ in this case,
produces the envelope at low equivalent widths. 
The dotted lines illustrate our completeness limits for
detection and are labeled with the fraction of galaxies detected along that locus.
The symbols denote the broadband \gi\ color of each galaxy (see key in figure).
}
\label{fig:on_l} \end{figure}

\begin{figure} \caption{Response of our filters to Coleman (1980)
spectral templates versus redshift.
The solid, dotted, and dashed lines represent the SED of 
an E/S0,  Sbc, and Im galaxy respectively.
{\it (Bottom)}
The on-band excess created for each spectrum as a function of
redshift.
{\it (top)}
The \gi\ color versus redshift.
Intergalactic attenuation is included.
}
\label{fig:ew2} \end{figure} 

\begin{figure} \caption{Equivalent measured from narrowband images (this paper)
versus equivalent width measured spectroscopically (Dressler \et 1999).}
\label{fig:ew_d99} \end{figure}

\begin{figure} \caption{Distribution of star formation rates.  Circles
represent our  \Ab851 results. Triangles show the field
at $z \sim 0.4$ (Hogg \et 1998).  The cluster and field data have
been converted to a SFR using $L_{\Ha}{\rm(observed)} \approx 1.17 L_{[OII]} 
{(\rm observed)}$ and the relation $\dot{M}_* = L_{\Ha} E_{\Ha} / 1.36 \times 
10^{41}$ergs/s.  The  \Ha luminosity is assumed to be attenuated by
a factor of 1.75, corresponding to $A_B = 1.0$ (Seaton 1979). For comparison,
the  local \Ha luminosity functions from Gallego \et and Tresse \et are shown.
All results are scaled to \h0 = 70~km/s/Mpc, $\Omega_M = 0.2$,
and $\Omega_{\lambda} = 0$.}
\label{fig:dsfr} \end{figure}

\begin{figure} \caption{Spatial distribution of objects with an on-band 
flux excess (top row) and an off-band flux excess (bottom row). The x and y
axes have tickmarks at 100 pixel intervals. The spatial distributions are
shown separately for galaxies with blue, green, and red \gi\ color.}
\label{fig:im_mo} \end{figure}

\begin{figure} \caption{Distribution of [OII] equivalent-widths for
the \Ab851 sample (solid bars) and the field sample (open bars).  
For cluster memebers, the rest-frame equivalent widths are a factor 
of $\sim 1.4$ lower.} 
\label{fig:ew} \end{figure}

\begin{figure} \caption{Observed equivalent width versus color relation
for the on-band-selected sample.  Galaxies with large color errors are shown 
by open squares at \gi = 0. The {\it X}s denote galaxies in which the 
on-band excess is not attributed to line emission.  The evolution of a galaxy 
is illustrated for four star formation histories.
A stellar population with a  constant SFR is shown at ages of
4~Myr, 40~Myr, 400~Myr, 1~Gyr, and 10~Gyr by 
from left to right by the solid squares.
The three sequences below illustrate a SFR that declines
by a factor of 0.37 over 10~Myr, 100~Myr, and 1~Gyr.
The stellar population along each fading track is shown at ages of  
$\tau$, 2$\tau$, 4$\tau$, and 6$\tau$ (from top to bottom). 
All models include an internal extinction of $A_B = 1.0$~magnitude.
}
\label{fig:ew_gi} \end{figure}

\begin{figure} \caption{
Color -- magntidue diagram for the on-band-selected galaxies.
Asterisks and circles denote the starburst population and
the [OII]-emitters with intermediate colors, respectively.
The fraction of interlopers without [OII] emission is likely
higher in the redder population represented by the X's.
The heavy lines outline the region populated by modeled stellar 
populations that contain $10^9$\msun\ of stars (see text for
details). Congruent locii are indicated for $10^8$\msun, 
$10^{10}$\msun, and $10^{11}$\msun populations.  The models have 
been reddened with a Galactic extinction law assuming $A_B = 1.0$~mag. 
}
\label{fig:mass} \end{figure}

\begin{figure} \caption{Remnants of the [OII]-selected galaxies in \Ab851 
at $z \sim 0$. A single-burst scenario is assumed for purposes of illustration.
{\it Top: } Number - absolute magnitude relation. 
The galaxies with  $\gi\ < 1.0$ and those with
$1.0 < \gi\ < 2.0$ populate the solid and open histograms respectively.
{\it Middle: } Same as top panel, but the histogram shading depicts
star formation timescale instead of color; from left to right $\tau$ equals
1~Gyr, 100~Myr, and 10~Myr. 
{\it Bottom: }  Duty-cycle correction. For this illustration, we assumed each 
galaxy bursts only once and that bursts occur at a constant rate in the 
cluster between redshifts $z \sim 1.0$ and $z \sim 0.4$. The dashed line 
illustrates the total number of remnants from the [OII]-selected population.
The solid line represents the magnitude distribution of galaxies classified 
as dwarf ellipticals in the VCC.  
}
\label{fig:fade_nm} \end{figure}

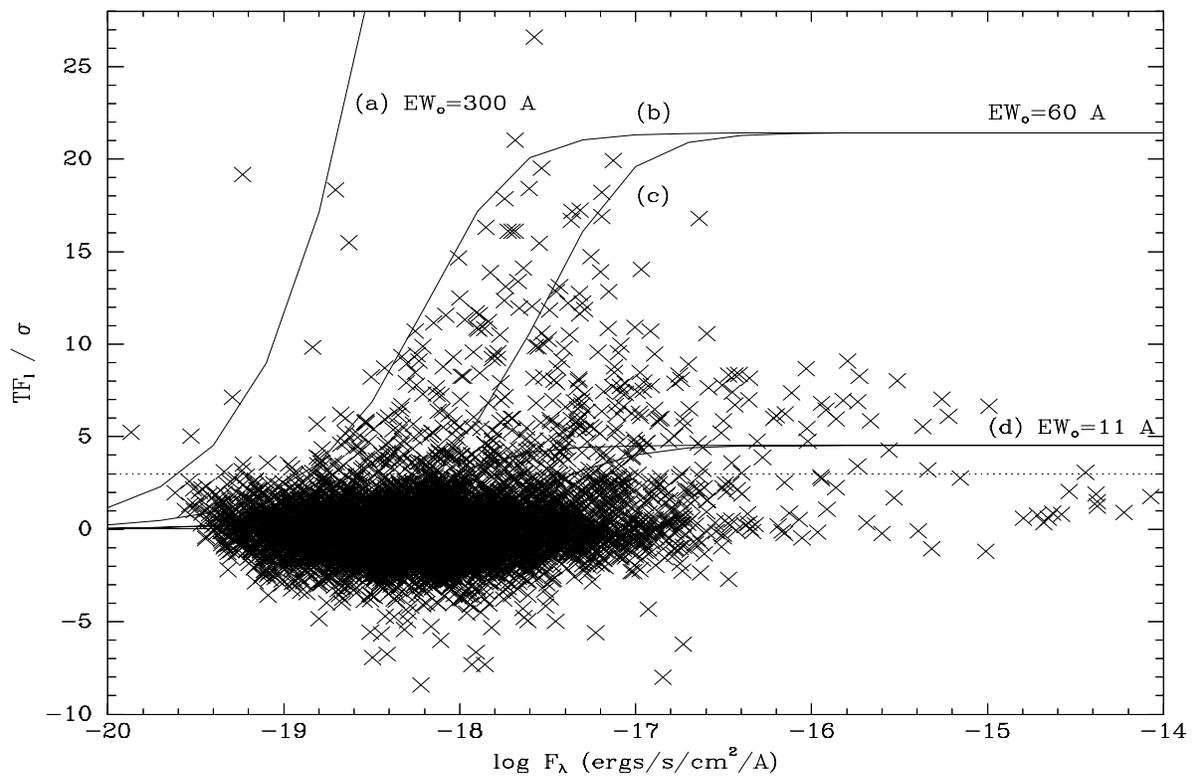
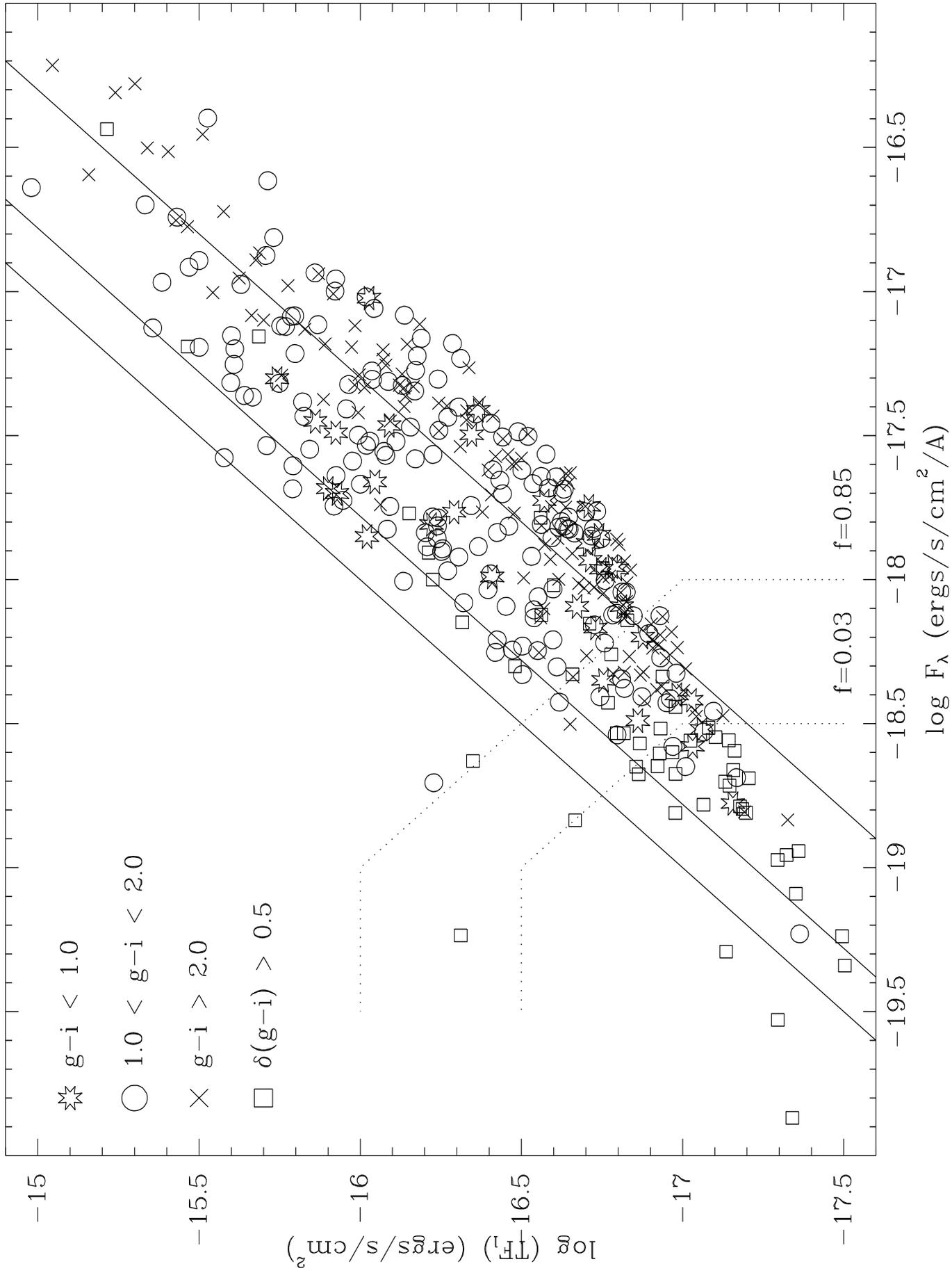
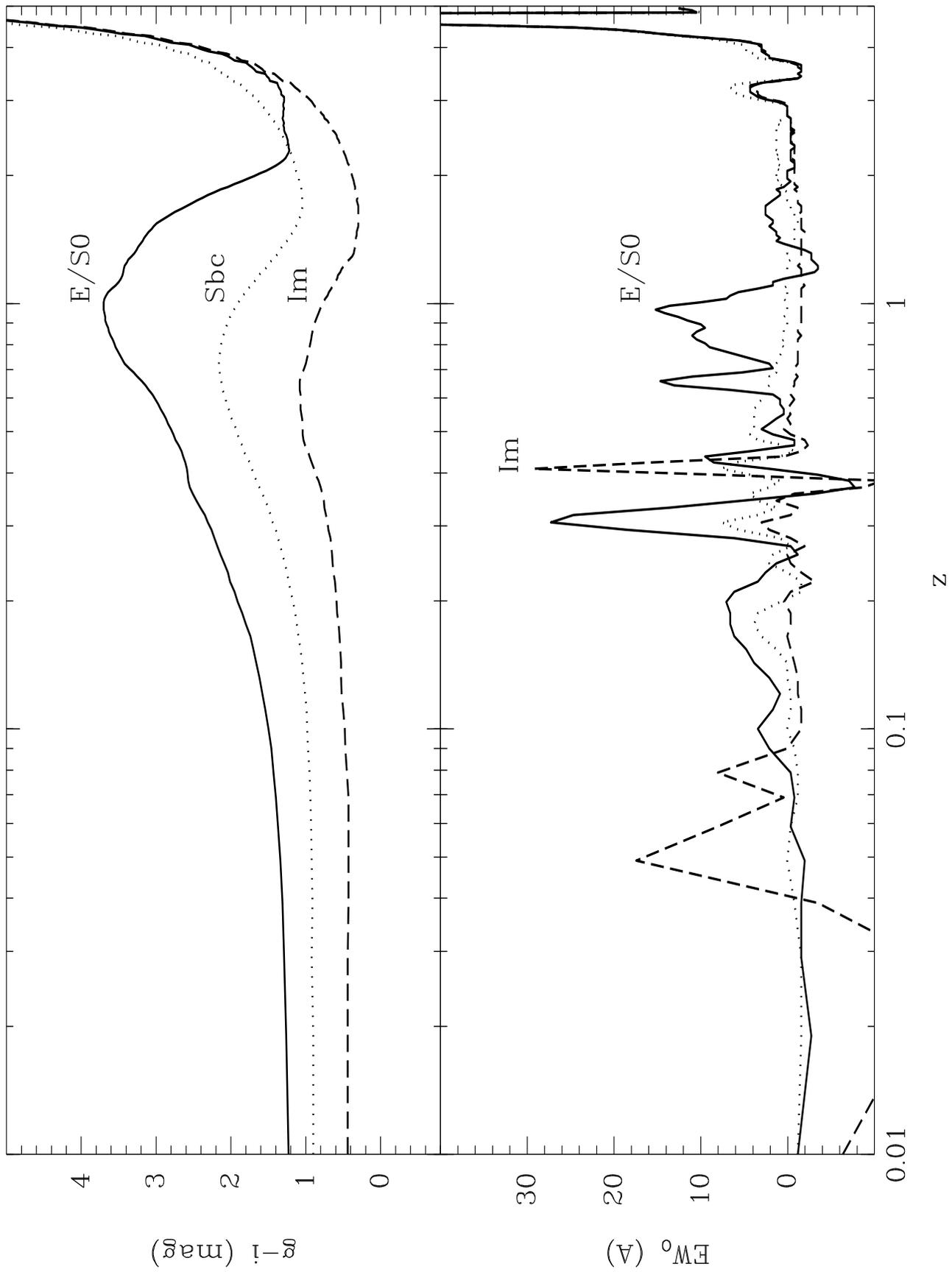
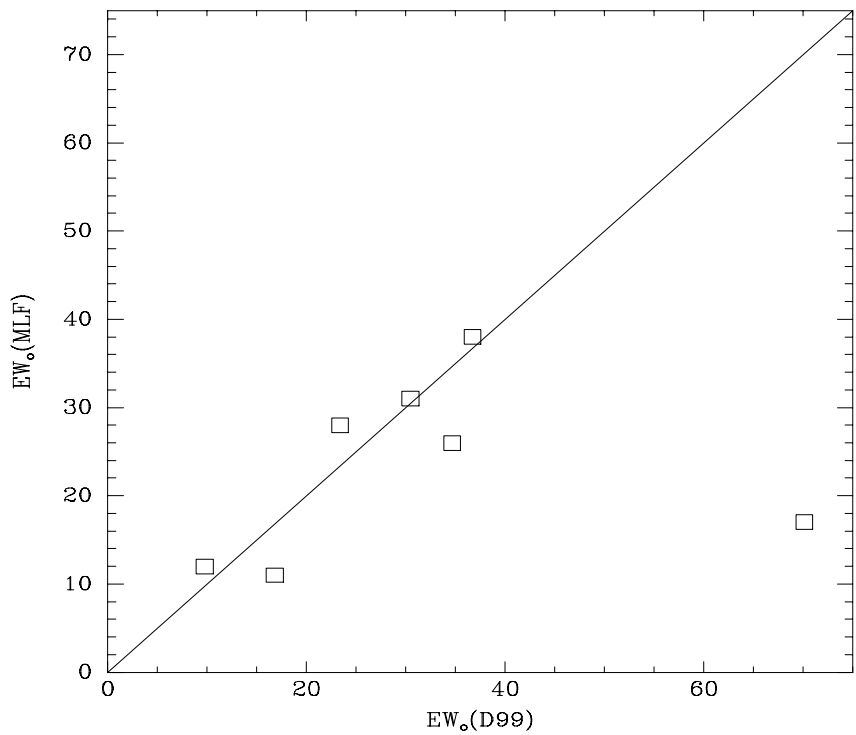
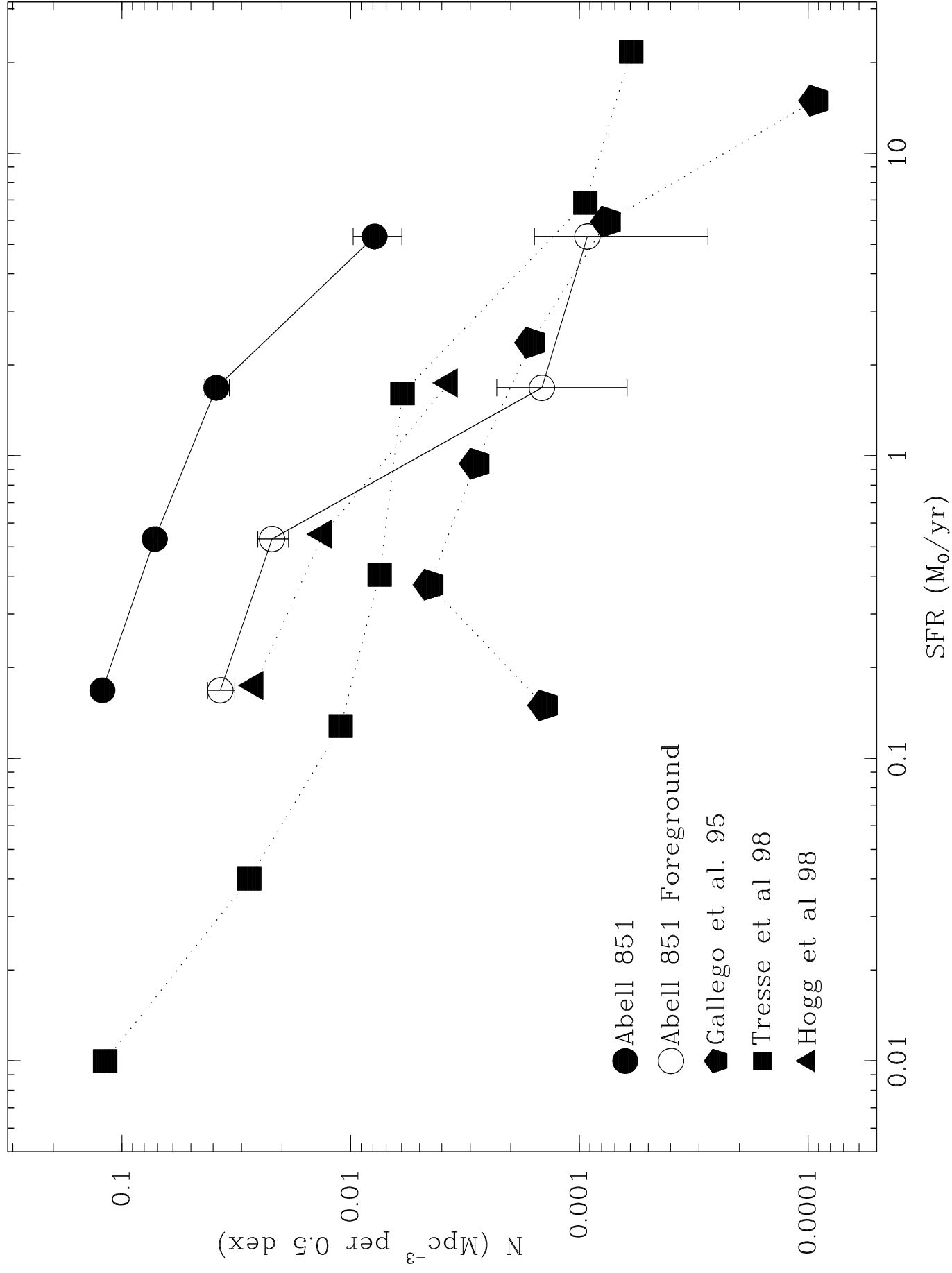
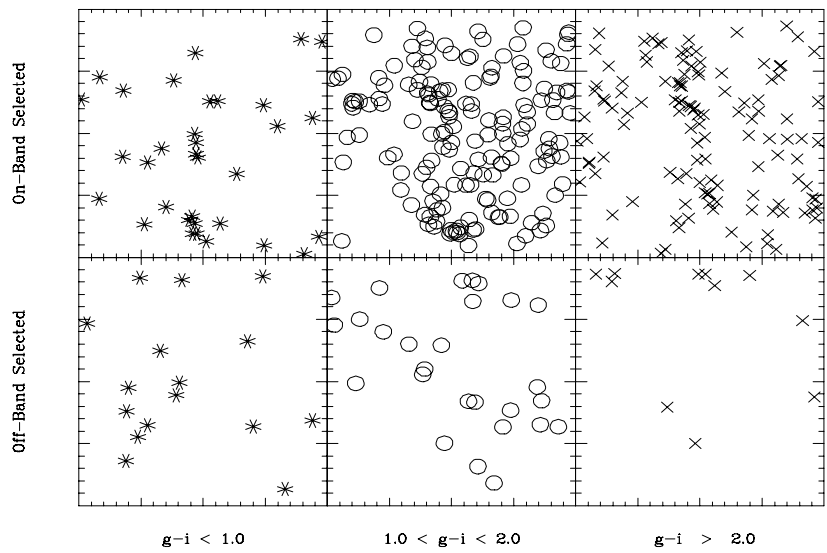
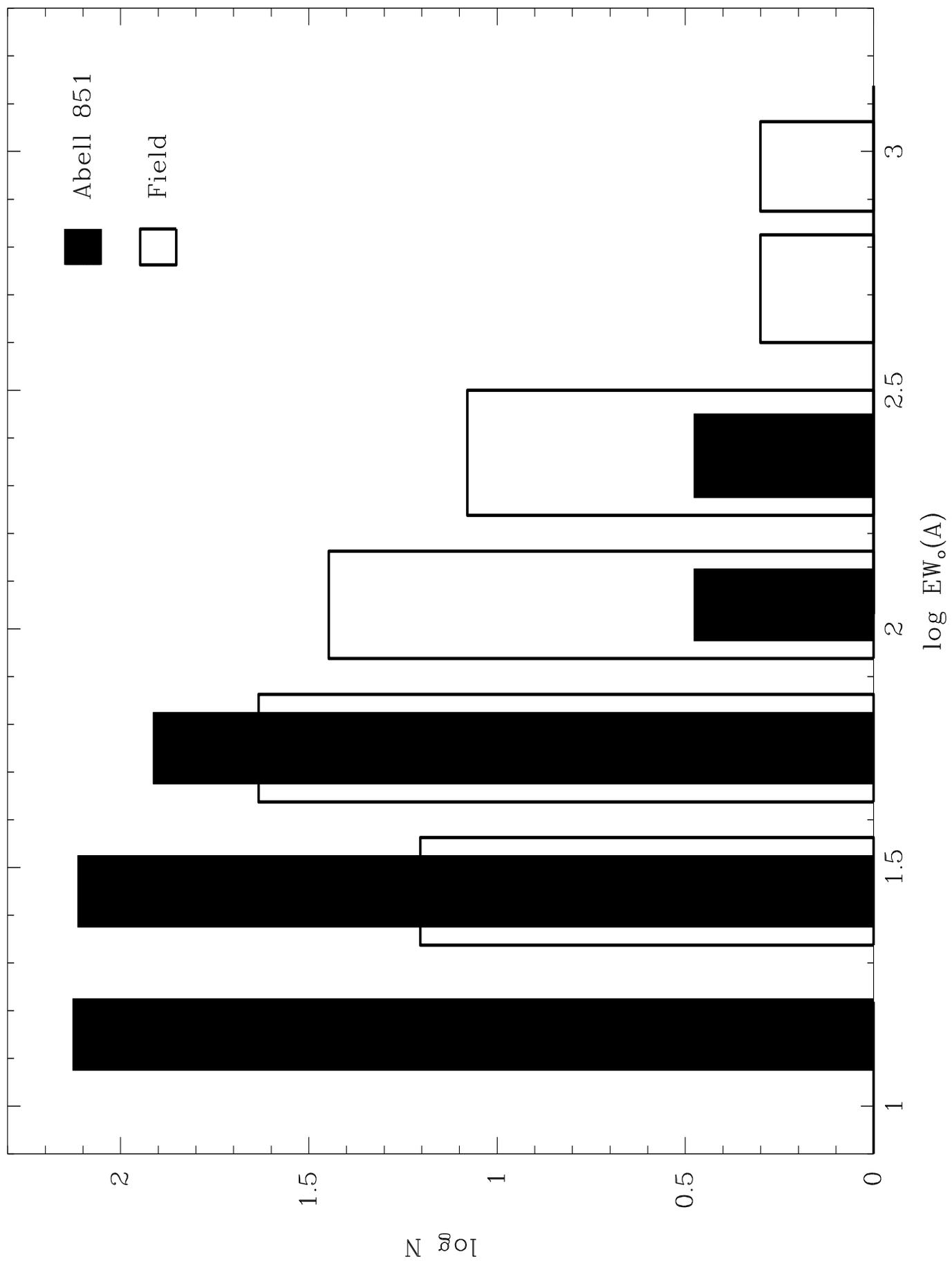
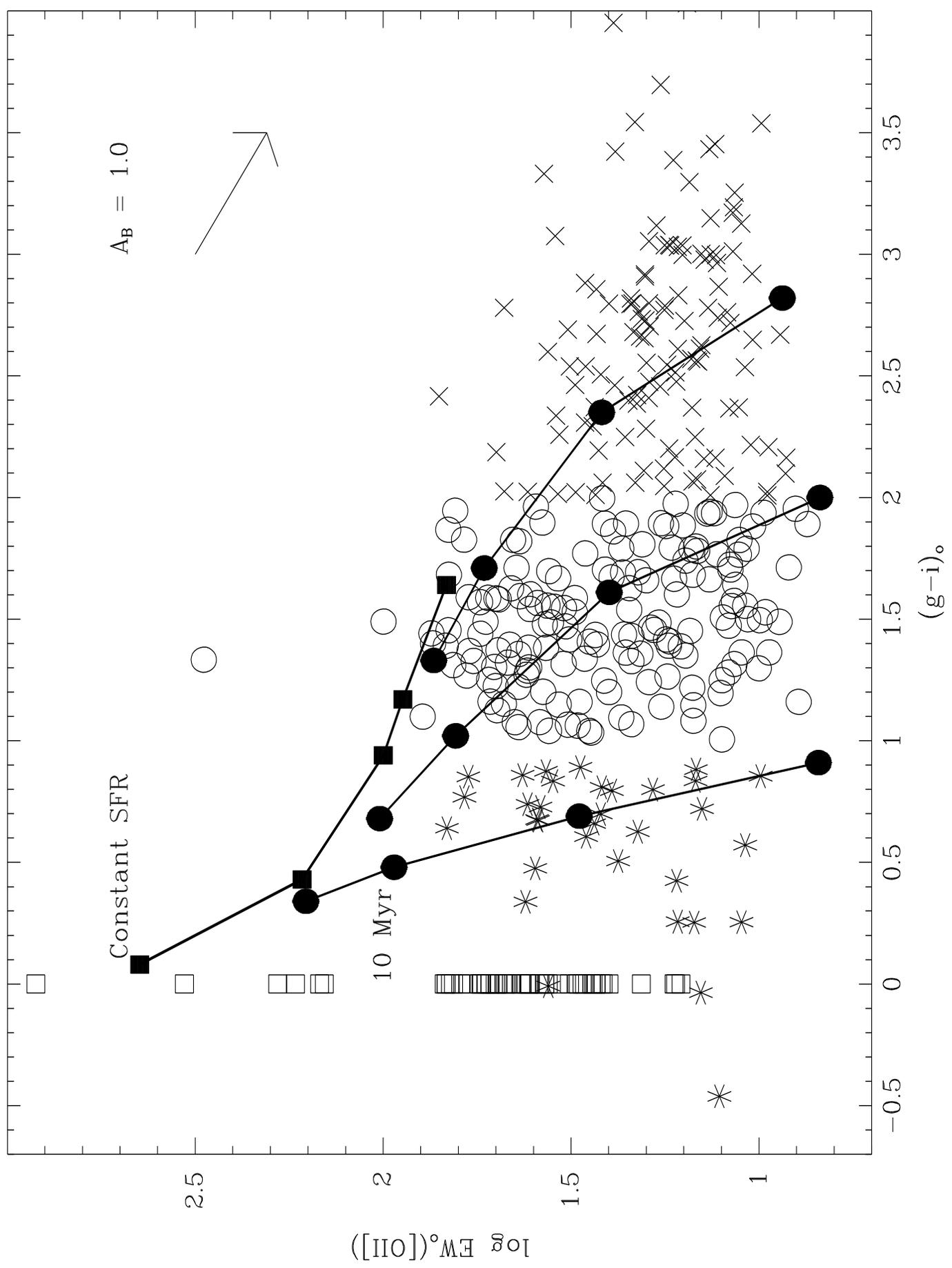
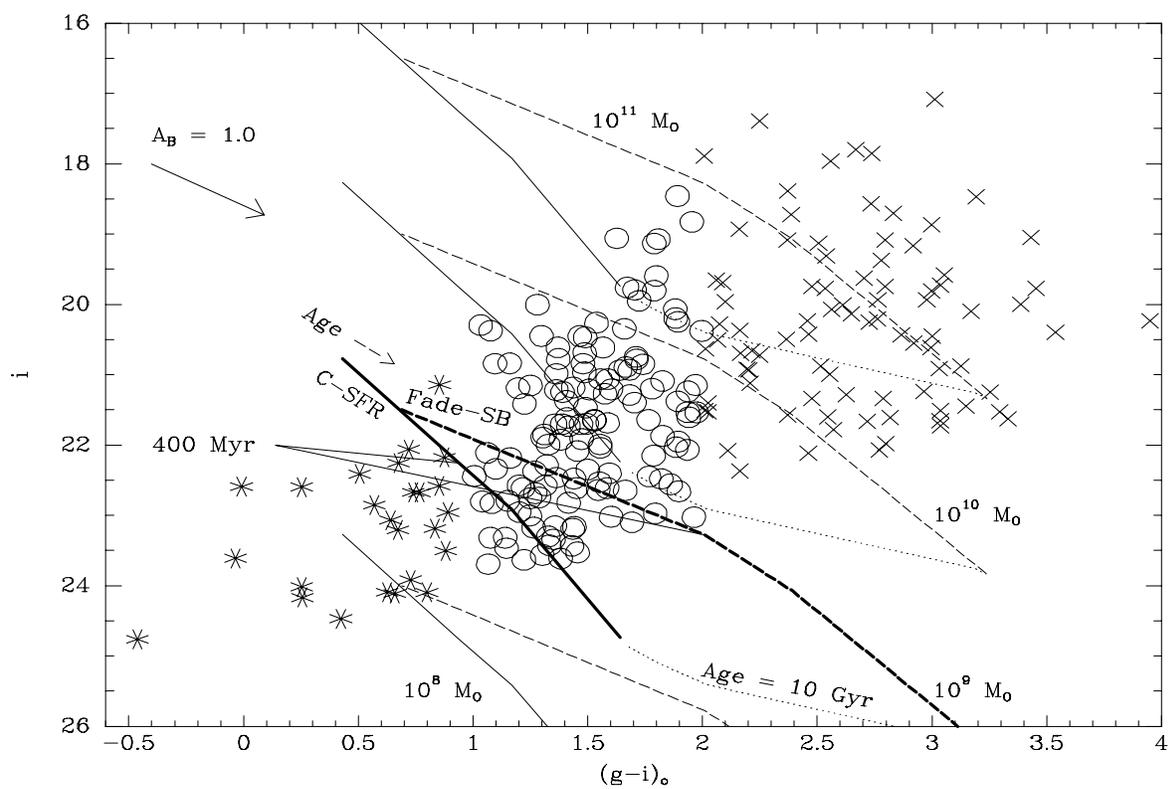
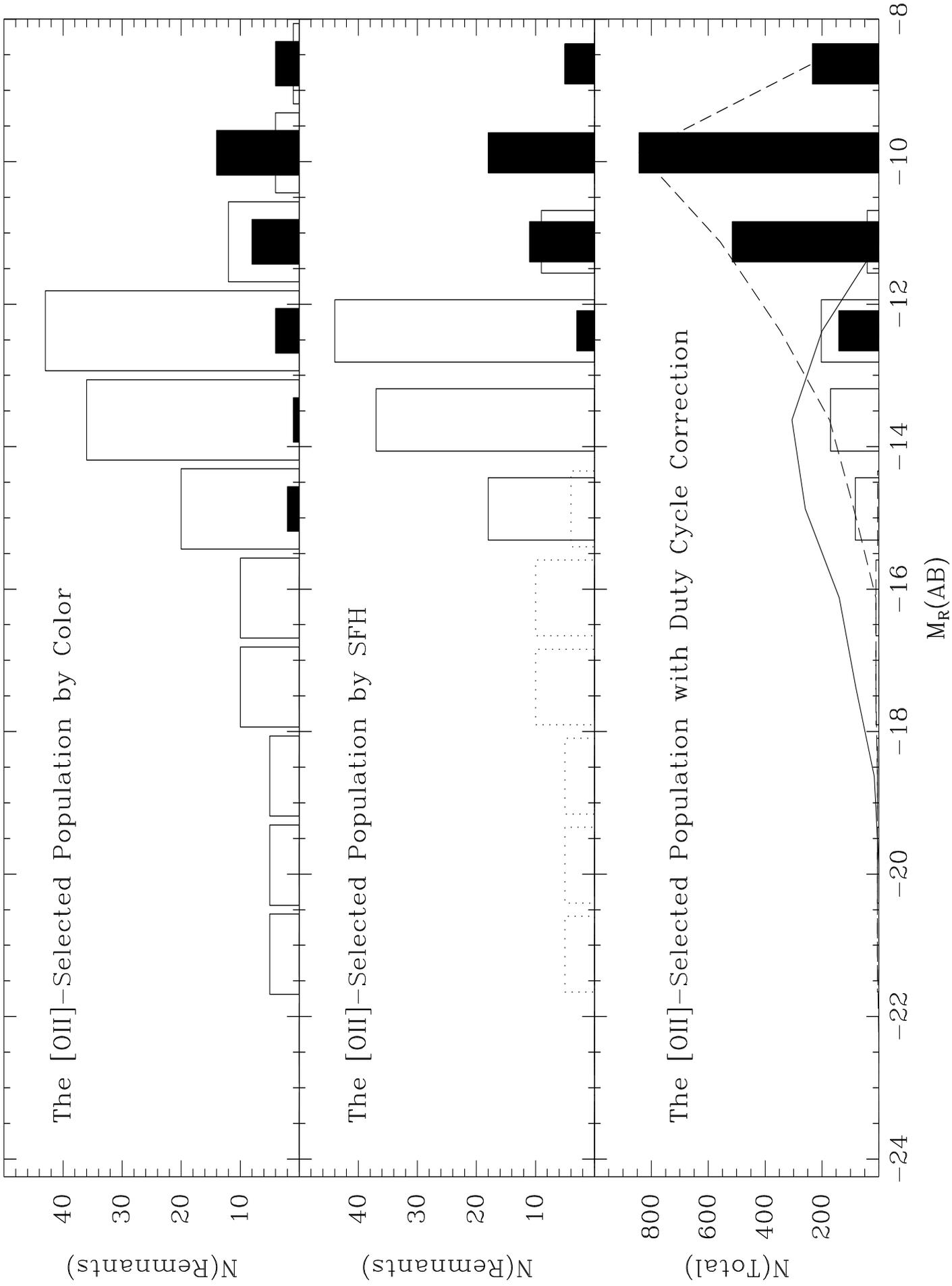
\end{document}